\documentclass[5p,times]{elsarticle}
\pdfoutput=1

\usepackage{lineno,hyperref}
\usepackage{amsmath}
\usepackage{graphicx}
\usepackage{caption}
\usepackage{subcaption}
\usepackage{verbatim} 
\usepackage{dingbat}
\graphicspath{ {images/} }

\newcommand{\mathbbm}[1]{\text{\usefont{U}{bbm}{m}{n}#1}} % from mathbbm.sty
\newcommand{\diag}{\mathop{\mathrm{diag}}}

\journal{Optics Communications}

\biboptions{sort&compress}
\begin{document}

\begin{frontmatter}

\title{Robust Two-Step phase estimation using the Simplified Lissajous Ellipse Fitting method with Gabor Filters Bank preprocessing}


\author[CIMAT,UPB]{V\'ictor H. Flores\corref{mycorrespondingauthor}}
\ead{victor.flores@cimat.mx}
\author[CIMAT]{Mariano Rivera}
\address[CIMAT]{Centro de Investigaci\'on en Matem\'aticas AC, 36023, Guanajuato, Gto., M\'exico.}
\address[UPB]{Departamento de Ingenier\'ia Rob\'otica, Universidad Polit\'ecnica del Bicentenario, 36283, Silao, Gto., M\'exico.}
\cortext[mycorrespondingauthor]{Corresponding author}

\begin{abstract}
We present the Simplified Lissajous Ellipse Fitting (SLEF) method for the calculation of the random phase step and the phase distribution from two phase-shifted interferograms. We consider interferograms with spatial and temporal dependency of background intensities, amplitude modulations and noise. Given these problems, the use of the Gabor Filters Bank (GFB) allows us to filter--out the noise, normalize the amplitude and eliminate the background. The normalized patterns permit to implement the SLEF algorithm, which is based on reducing the number of estimated coefficients of the ellipse equation, from five terms to only two. Our method consists of three stages. First, we preprocess the interferograms with GFB methodology in order to normalize the fringe patterns. Second, we calculate the phase step by using the proposed SLEF technique and third, we estimate the phase distribution using a two--steps formula. For the calculation of the phase step, we present two alternatives: the use of the Least Squares (LS) method to approximate the values of the coefficients and, in order to improve the LS estimation, a robust estimation based on the Leclerc's potential. The SLEF method's performance is evaluated through synthetic and experimental data to demonstrate its feasibility.
\end{abstract}


\end{frontmatter}

%\linenumbers

%-------------------------------------------------------------------------------------------------
\section{Introduction}
\label{sec:intro}
%-------------------------------------------------------------------------------------------------

\noindent Phase shifting interferometry is widely used to obtain the phase distribution in interferometric measurements \cite{malacara2007optical, servin2014fringe, creath1993temporal}. Even though the calculation can be performed in a single shot \cite{takeda1982fourier}, the use of several phase shifted interferograms proved to make the measurement more robust to environmental variations \cite{carre1966installation}. Nowadays, the tendency has been to reduce the number of steps in order to measure dynamic events \cite{rodriguez2008one, toto2014dynamic, munoz2015measurement, kimbrough2006pixelated, meng2006two, vargas2011two, kreis1992fourier}.

One of the main challenges in interferometry is the variation of the parameters of the intensity map. The spatial and temporal dependency of the background intensity, the amplitude modulation and noise are common in non-aligned arrangements \cite{malacara2007optical}. These issues also apply to one-shot interferometry, where the use of optical components such as diffractive devices, polarizers or pixelated masks disturb the captured interferograms \cite{rodriguez2008one, toto2014dynamic, munoz2015measurement, kimbrough2006pixelated}.

Mathematically, the intensity model of these variable phase--shifted interferograms is given by

\begin{equation} \label{eq:image}
	I_k(p) = a_k(p) + b_k(p)\cos[\phi(p) + \delta_k] + \eta_k(p),
\end{equation}
where $k \in 1,2$ is the interferogram index, $p$ is pixel's coordinates in the regular lattice $\mathcal{L}$, $a$ is the background component, $b$  is the fringe's amplitude function, $\phi$ is the phase to be recovered, $\delta_k$ is the random phase step and $\eta_k$ is additive noise. For the case of  two--step algorithms, we can assume that $\delta_1 = 0$ and $\delta_2 = \delta$. The $a$ and $b$ dependencies on $k$ do not allow one to use the known algorithms of phase extraction since they assume temporally constant the background intensity and the amplitude modulation. In such cases, a preprocess is needed in order to normalize the patterns and compute the phase.

In this paper we propose a novel, simplified and more robust version of the Lissajous Ellipse Fitting (LEF) method proposed in \cite{farrell1992phase} for estimating the arbitrary phase step between two phase--shifted fringe patterns with variable parameters. As mentioned before, due to the tendency of reducing the number of steps to estimate the phase distribution, the two--step algorithms have attracted considerably the attention in the past few years; for these reason, several algorithms have been proposed such as Refs.  \cite{meng2006two, vargas2011two, kreis1992fourier,  vargas2012two, deng2012two, farrell1992phase, rivera2016two, trusiak2015two, meng2008wavefront, wielgus2015two, kulkarni2018two, liu2015phase, saide2017evaluation}. Particularly, various algorithms based on the LEF method have been designed to improve the estimation of the phase step and the phase distribution in two--step intertferometry, such as: iterative processes based on the least square technique \cite{zhang2018random}, the use of the Gram-Schmidt orthonormalization to transforms the ellipse into a circle \cite{zhang2019two}, the application of a Hilbert-Huang Transform (HHT) pre--filtering with the LEF algorithm \cite{liu2016simultaneous} or the computation of the Euclidean distance from the points to the ellipse \cite{meneses2015phase}; just to mention some of the novel techniques. 

We named our proposed method as Simplified Lissajous Ellipse Fitting (SLEF). Our algorithm reduces the number of estimated coefficients of the ellipse equation, from five terms to only two. Consequently, it is improved the accuracy on the estimation of the relevant parameters by reducing the overfitting of the ellipse to residual noise. Our method consists of three stages: First, we preprocess the fringe patterns using a Gabor Filter Bank (GFB) in order to remove the background variation, normalize the amplitude modulation and filter--out noise  \cite{rivera2016two}. Second, we calculate the phase step through the two term expression of the ellipse equation by using the Least Square (LS) method; alternatively,  by minimizing a cost function based on the Leclerc's potential to define a Robust Estimator (RE) of the coefficients. Third, we calculate the phase distribution using the two--steps algorithm reported in Ref. \cite{muravsky2011two}. 

We will demonstrate that the use of a GFB as a filtering preprocess does not only improves the robustness of the phase extraction, but also, as it will be demonstrated in Section \ref{sec:SLEF}, it simplifies the computation of the ellipse coefficients and consequently the phase step estimation. We remark that we can replace the GFB based preprocessing with other normalization techniques that provide the elimination of the background component, normalize the amplitude modulation and filter--out the noise; for example the Windowed Fourier Transform \cite{kemao2007two},  the HHT \cite{trusiak2014advanced}  or isotropic normalization \cite{quiroga2003isotropic} among others.

Our main contributions are:
\begin{enumerate}
	\item A method that uses only 2 parameters that produce equivalent results as the more complex method that uses 5 parameters. 
	\item A robust estimation which allows us to overcome large residuals of the pre--filtering process.
	\item We demonstrate by numerical experiments that the LEF methods are more accurate with normalized patterns with GFB than the HHT.
\end{enumerate}

%-------------------------------------------------------------------------------------------------
\section{Related Methods}
In this section, we will present a brief review of the methods to be used in our proposal (which is describe in Section \ref{sec:SLEF}): The Lissajous Ellipse Fitting and the Gabor Filters Bank.

\subsection{Brief review of the Lissajous Ellipse Fitting (LEF) method}
\label{sec:LEF}
%-------------------------------------------------------------------------------------------------

\noindent The Lissajous Ellipse Fitting (LEF) method consists on using the Lissajous figure to detect the phase step between two interferograms and estimate the phase \cite{farrell1992phase, liu2016simultaneous, zhang2018random, zhang2019two, meneses2015phase}. 

\begin{figure}[ht]
	\centering
	\includegraphics[width=.32\linewidth]{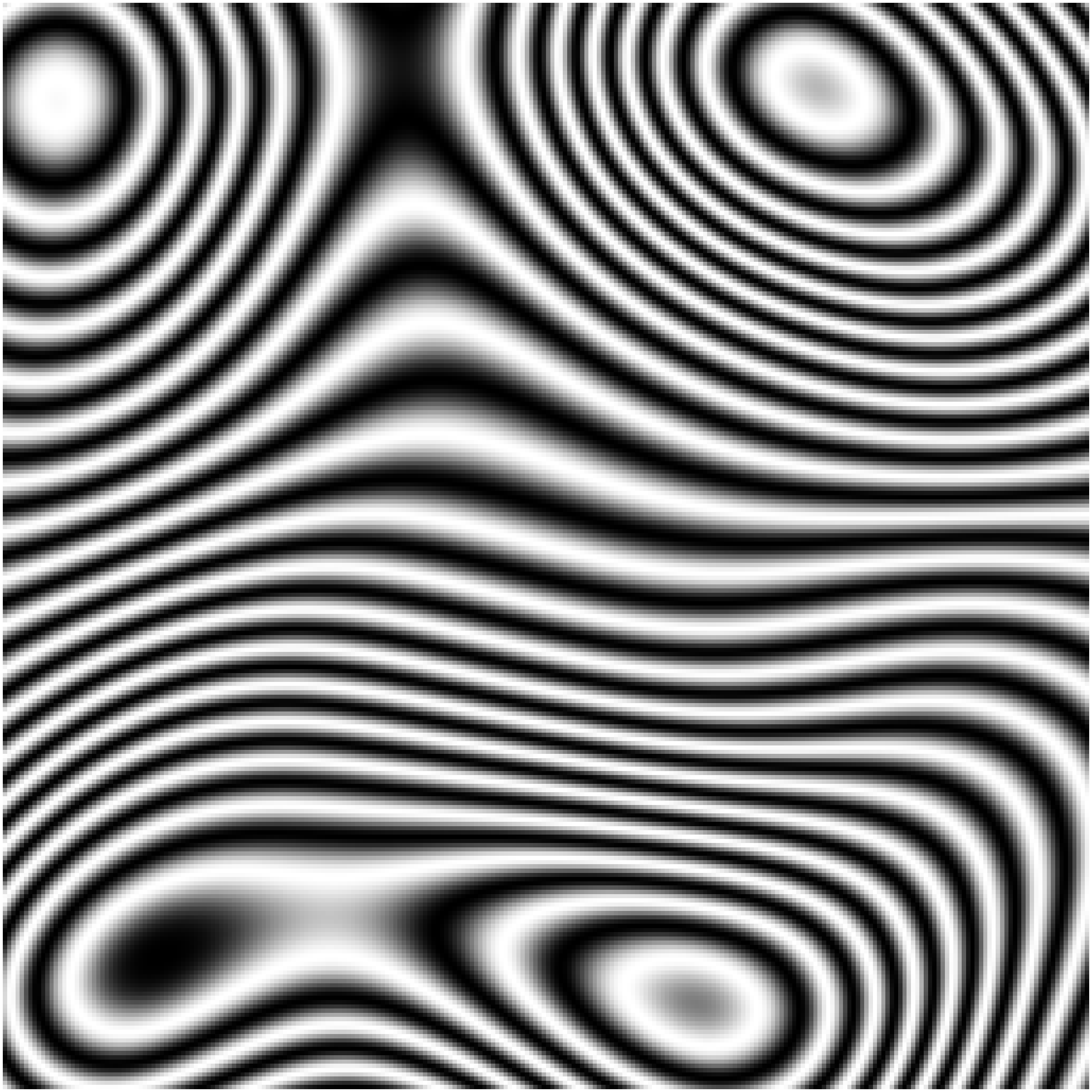}
	\includegraphics[width=.32\linewidth]{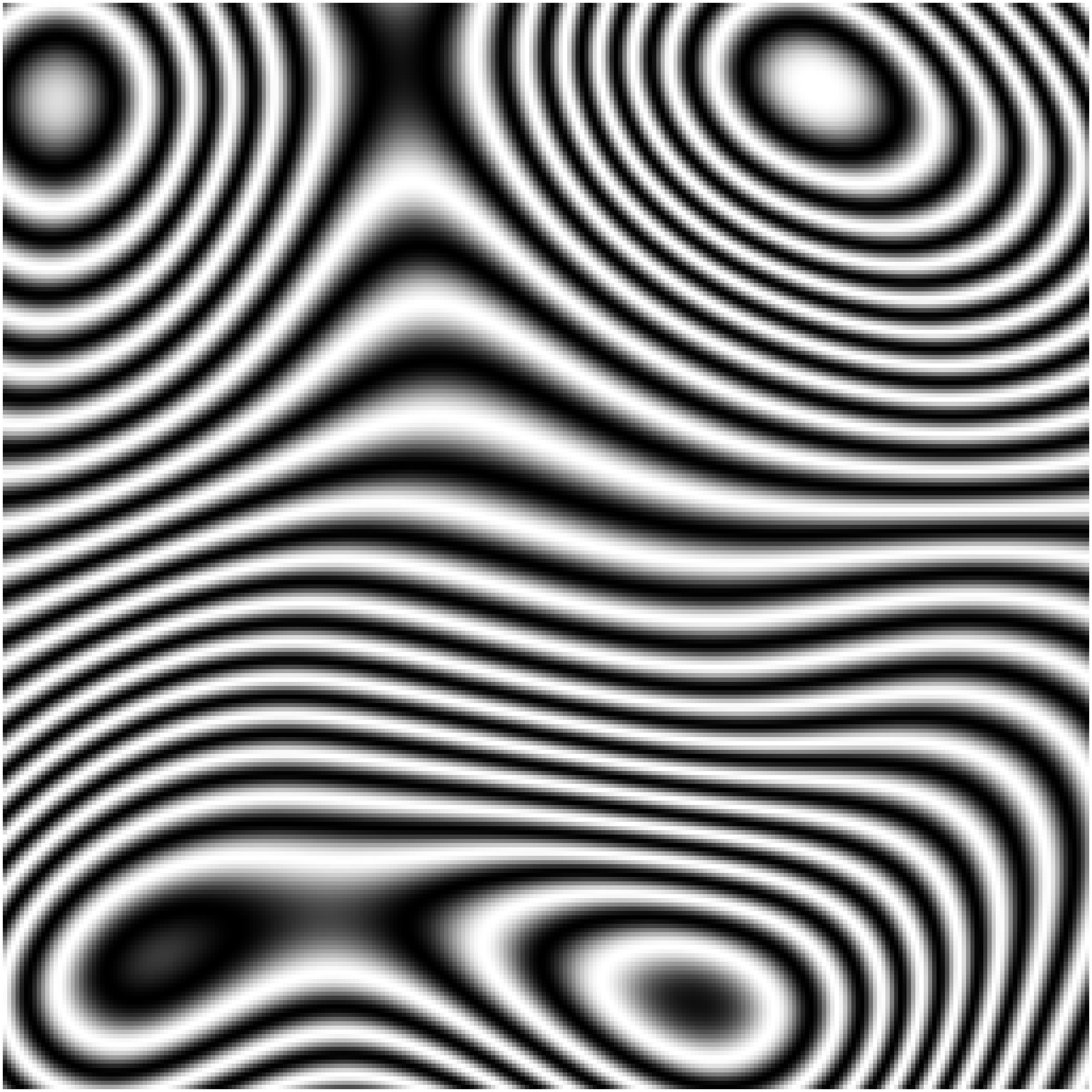}
	\includegraphics[width=.32\linewidth]{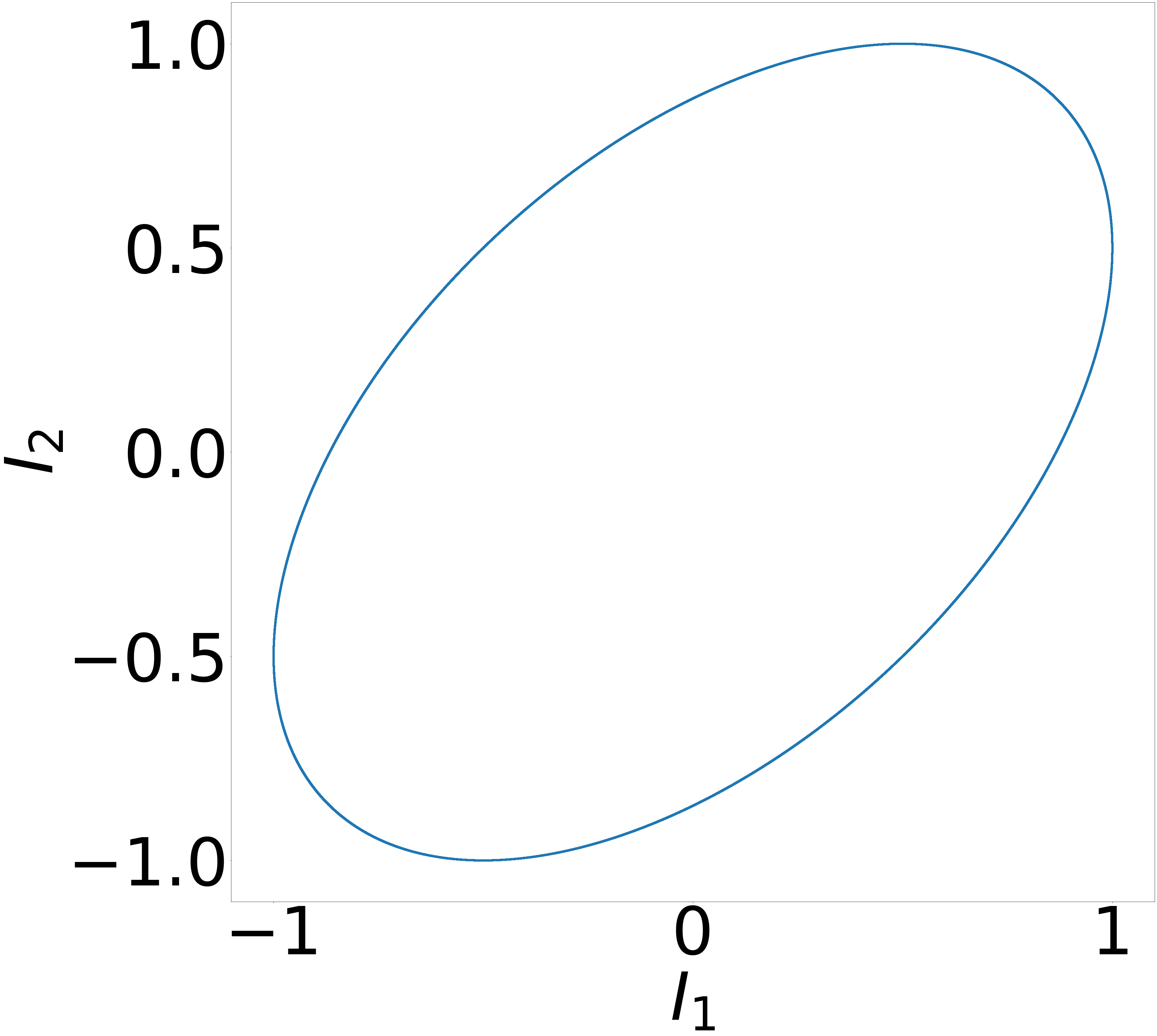}
	\caption{Resulting Lissajous ellipse of mapping pixel-wise the interferograms corresponding intensities of two ideal interferograms.}
	\label{fig:lef}
\end{figure}

Two phase-shifted interferograms can be represented as the Lissajous figure by plotting their pixel--wise corresponding  intensities, see Figure \ref{fig:lef}. The relation between the major and the minor axes is the result of the phase--shift \cite{farrell1992phase}: if $\delta = \pi/2$, the ellipse would become a circle.

For the case of two-step interferometry, one can consider that the background intensity and the amplitude term are spatially constant and timely invariant, , $a_1(p) = a_2(p) = a$ and $b_1(p) = b_2(p) = b$,  and that the noise $\eta_k(p)$ is filtered--out. Then, by performing the addition and subtraction of these interferograms, one obtains:

\begin{equation} \label{eq:add}
	I_{add} = I_1 + I_2 = 2a + 2b \cos\left(\phi + \frac{\delta}{2}\right) \cos\left(\frac{\delta}{2}\right)
\end{equation}
\begin{equation} \label{eq:sub}
	I_{sub} = I_1 - I_2 = 2b \sin\left(\phi + \frac{\delta}{2}\right) \sin\left(\frac{\delta}{2}\right),
\end{equation}
where the spatial dependency of $\phi$ is omitted in order to simplify the notation.

By solving equations \eqref{eq:add} and \eqref{eq:sub} for $\cos(\phi + \delta/2)$ and $\sin(\phi + \delta/2)$ respectively, and considering that $\cos^2(z) + \sin^2(z) = 1$, one obtains the expression of an ellipse represented as:

\begin{equation} \label{eq:eli1}
	\left(\frac{I_{add} - x_0}{\alpha_x}\right)^2 + \left(\frac{I_{sub} - y_0}{\alpha_y}\right)^2 = 1
\end{equation}
where $x_0 = 2a, y_0 = 0, \alpha_x = 2b\cos(\delta/2)$ and $\alpha_y = 2b\sin(\delta/2)$.

%By expanding equation \eqref{eq:eli1} as
%
%\begin{equation} \label{eq:eli2}
%	\frac{1}{\alpha_x^2} I_{add}^2 + \frac{1}{\alpha_y^2} I_{sub}^2 - \frac{2x_0}{\alpha_x^2} I_{add} - \frac{2y_0}{\alpha_y^2} I_{sub} + \frac{x_0^2}{\alpha_x^2} + \frac{y_0^2}{\alpha_y^2} -1 = 0
%\end{equation}
Then, one can rewrite \eqref{eq:eli1} in the conical equation of the ellipse:

\begin{equation} \label{eq:eli3}
	\theta_1x^2 + \theta_2y^2 + \theta_3x + \theta_4y + \theta_5 = 0
\end{equation}
where $\theta_1 = \frac{1}{\alpha_x^2}, \theta_2 = \frac{1}{\alpha_y^2}, \theta_3 = - \frac{2x_0}{\alpha_x^2}, \theta_4 = - \frac{2y_0}{\alpha_y^2}$ and $\theta_5 =  \frac{x_0^2}{\alpha_x^2} + \frac{y_0^2}{\alpha_y^2} -1$. Thus, by solving the coefficients (vector $\theta$) by the least square method (as proposed in \cite{farrell1992phase, liu2016simultaneous, zhang2018random, zhang2019two, meneses2015phase}), the phase step is computed as

\begin{equation} \label{eq:delta} 
	\delta = 2\arctan\left(\sqrt{\frac{\theta_1}{\theta_2}}\right).
\end{equation}
Hence, the phase distribution is calculated with

\begin{equation} \label{eq:phi1} 
	\phi = \arctan \left( \frac{I_{sub}}{I_{add} + \frac{\theta_3}{2\theta_1}} \sqrt{\frac{\theta_2}{\theta_1}}\right) - \frac{\delta}{2}.
\end{equation}

In this work, we present a simplified extension of the LEF method and demonstrate its reliability with complex fringe pattern sets.

%-------------------------------------------------------------------------------------------------
\subsection{Brief review of Gabor Filters Bank (GFB)}
\label{sec:GFB}
%-------------------------------------------------------------------------------------------------
As described by \cite{daugman1985uncertainty, daugman1988complete, daugman1993high, rivera2016two, rivera2018robust, jun2002strain}, a Gabor Filter ($GF$) is a complex band--pass filter created from the modulation of a complex sinusoidal function with a Gaussian filter ($G$). The complex response from this filter is modeled as:

\begin{equation} \label{eq:GF}
	GF\{I\}(x,\omega) = I(x)\otimes[e^{-i\omega x}G(x,\sigma)]
\end{equation}
where $I$ is the image to be filtered, in this case the fringe patterns, $x$ is the pixel's index,  $\omega$ is the tuned frequency of the filter, $\sigma$ is the Gaussian filter width (window size) and $\otimes$ denotes the convolution of the functions. In terms of frequency, the window size $\sigma$ of the filter represents the width of the band--pass filter centered at the $\omega$ frequency.

A GFB is a set of GFs defined by a set of frequencies $\{\omega_k\}_{k=1,2,\ldots}$ and windows sizes $\{\sigma_k\}_{k=1,2,\ldots}$. One of the filtered images of the bank would be given as

\begin{equation} \label{eq:GF2}
	\tilde{I}_k(x) = GF\{I\}(x,\omega_k) = I(x)\otimes[e^{-i\omega_k x}G(x,\sigma_k)],
\end{equation}
which also can be expressed as

\begin{equation} \label{eq:GF3}
	\tilde{I}_k(x) = m_k(x)e^{-i\psi_k(x)}
\end{equation}
where $m_k$ is the magnitude and $\psi_k$ is the phase of the response of the image to the $k^{th}$ filter. Then, the filter with the maximum response is estimated at each $x$ index as

\begin{equation} \label{eq:GF4}
	k^*(x) = \underset{k}{\mathrm{argmax}} \,m_k(x).
\end{equation}

Hence, the local magnitude and phase of the filtered pattern would be 
\begin{align}\label{eq:GF5}
	m(x) = m_{k^*(x)}(x) \\
	\psi(x) = \psi_{k^*(x)}(x). 
\end{align}

Finally, the normalized and filtered fringe pattern is given by 

\begin{equation} \label{eq:GFB}
	\hat{I}(x) = \cos[\psi(x)].
\end{equation}
In fact, one could say that trough this method the phase of the fringe pattern is recovered, which is the goal, but the main issue is presented with closed fringes where the sign ambiguity is cannot be solved with only one fringe pattern.

\begin{figure}[ht]
	\centering
	\includegraphics[width=.5\linewidth]{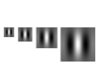}
	\caption{Real component of the GFs corresponding to the first orientations and periods equal to $[7,10,15,25]$ pixels, respectively.}
	\label{fig:real_gfb}
\end{figure}

In our synthetic experiments we use images of  $256 \times 256$ pixels. We use a GFB  with ten orientations ($\theta_k = k*\pi/10$ for $k=0,2, \dots,9$), four frequencies corresponding to the periods of pixels $\tau = [7,10,15,25]$, a Gaussian window with standard deviation  equal the half of the filter period and a window size equal the double of the period. Fig. \ref{fig:real_gfb} depicts the real component of the GFs corresponding to the first orientation.

%-------------------------------------------------------------------------------------------------
\section{Simplified Lissajous Ellipse Fitting (SLEF) Method}
\label{sec:SLEF}
%-------------------------------------------------------------------------------------------------

%-------------------------------------------------------------------------------------------------
\subsection{Least Squares based SLEF }
\label{ssec:SLEF-LS}
%-------------------------------------------------------------------------------------------------
\noindent Herein we introduce our extension to the LEF algorithm for estimating the actual phase step. We named our variant as SLEF. For this purpose, we will consider the intensity model of the interferograms expressed in \eqref{eq:image}.

Considering the implementation of the GFBs (explained in section \ref{sec:GFB}), we eliminate the background variation, normalize the amplitude modulation and remove the noise (presented in \eqref{eq:image}). The ideally normalized two-step interferograms are:
\begin{align}\label{eq:imagenorm}
	\hat{I}_1(p) & = \cos[\phi(p)] \\ 
	\hat{I}_2(p) & = \cos[\phi(p) + \delta],
\end{align}
where by effect of the normalization we have $a_1(p) = a_2(p) = 0$ because of the background elimination, $b_1(p) = b_2(p) = 1$ because of the amplitude normalization and $\eta_1(p) = \eta_2(p) = 0$ because of the filtering process (similar to the one presented in \eqref{eq:GFB}). For the two--step algorithm, $\delta_1 = 0$ and $\delta_2 = \delta$.

Since we assume removed the background illumination variations, the center of the ellipse is at the origin because  $x_0 = 2a = 0$ and the eccentricity terms are given by 
\begin{align}\label{eq:consider} 
	\alpha_x&=2\cos\left(\frac{\delta}{2}\right)\\
	\alpha_y&=2\sin\left(\frac{\delta}{2}\right).
\end{align}

Thus, we can simplify the \eqref{eq:eli1} of the ellipse for the Lissajous pattern as

\begin{equation} \label{eq:eli1n}
	\left(\frac{\hat{I}_{add} }{\alpha_x}\right)^2 + \left(\frac{\hat{I}_{sub} }{\alpha_y}\right)^2 = 1,
\end{equation}
which corresponds to the equation of the ellipse centered at the origin. Hence, the ellipse's conical expression is given by 

\begin{equation} \label{eq:eli3n}
	\theta_1{\hat{I}_{add}} ^2 + \theta_2 {\hat{I}_{sub} }^2  -1 = 0.
\end{equation}

Note that in this ideal case, \eqref{eq:eli3n} is fulfilled for all the pixels. In case that the pre--filtering process does not guarantee an effective normalization of the patterns, the scatter set of points $(\hat{I}_{add}, \hat{I}_{sub})$ lay in an uncentered ellipse. Thus, we can center the points with

\begin{align}\label{eq:consider2} 
	x=\hat{I}_{add} - \langle\hat{I}_{add}\rangle \\
	y=\hat{I}_{sub} - \langle\hat{I}_{sub}\rangle
\end{align}
where $\langle\cdot\rangle$ denotes the operator that computes the mean. In Figure \ref{fig:filt_gfb} we present a sample of the effect of the filtering process on noisy and non-normalized patterns. The Lissajous ellipse, the line in red, can be considered as the mean of the observed values. Moreover, the spread points around the ideal ellipse are associated to the residual noise ($\varepsilon_1$ and $\varepsilon_2$) of the pre--filtering process. For this reason, we model such variations with an $\varepsilon$ term in \eqref{eq:eli3n}: 

\begin{equation} \label{eq:eli4n}
	\theta_1x^2 + \theta_2y^2  -1 = \varepsilon(x,y).
\end{equation}

Our SLEF method is based in this simplified equation with only two free parameters, $\theta_1$ and $\theta_2$ and the use of a GFB for normalizing and pre--filtering the FPs. Then, because the presence of the residual $\varepsilon$, we choose to solve the  overdetermined system \eqref{eq:eli4n} with the Least Square (LS) method 

%\begin{equation} \label{eq:lsqn}
%	\begin{bmatrix}
%   	 	\hat{I}_{add}(x_1)^2	&\hat{I}_{sub}(x_1)^2  &1 \\
%    		\hat{I}_{add}(x_2)^2	&\hat{I}_{sub}(x_2)^2  &1\\
%    		\vdots & \vdots \\
%    		\hat{I}_{add}(x_N)^2	&\hat{I}_{sub}(x_N)^2 &1
%	\end{bmatrix}
%	\begin{bmatrix}
%    		\theta_1 \\
%    		\theta_2 \\
%		1
%	\end{bmatrix}
%		=
%	\begin{bmatrix}
%    		1 \\
%    		1 \\
%    		\vdots \\
%    		1
%	\end{bmatrix}
%\end{equation}

\begin{equation} \label{eq:lsqn}
	\underset{\theta_1,\theta_2}{\arg\min}\frac{1}{2}\sum_{p \in \mathcal{L}} (\theta_1 x(p)^2 + \theta_2 y(p)^2 - 1)^2.
\end{equation}
This expression can be rewritten as

\begin{equation}\label{eq:lsqn2}
	\underset{T}{\arg\min}\frac{1}{2}||X^T T - \mathbbm{1}||_2^2
\end{equation}
where, in order to simplify the notation, we define $x_i = x(p_i)$ and $y_i = y(p_i)$. Thus
\begin{subequations}
	\begin{align}
		T&\overset{def}{=}[\theta_1, \theta_2]^T \\
		X^T&\overset{def}{=}
		\begin{bmatrix}
   	 		x_1^2	&y_1^2 \\
    			x_2^2	&y_2^2 \\
    			\vdots & \vdots \\
    			x_N ^2&y_N^2
		\end{bmatrix}
	\end{align}
\end{subequations}
and  $N = \sharp\mathcal{L}$, $\mathbbm{1}$ is a vector of $N$  entries equal $1$. Thus, the solution to \eqref{eq:lsqn2} is given by

\begin{equation}\label{eq:TLS}
	T = (XX^T)^{-1} X\mathbbm{1}.
\end{equation}
 
Once we have solved system \eqref{eq:lsqn2} for $\theta_1$ and $\theta_2$, we are in condition to calculate the phase step $\delta$ with \eqref{eq:delta}. Finally, we compute the phase distribution with the formula for two--step phase shifting reported in Ref. \cite{muravsky2011two}:

\begin{equation}\label{eq:phi2}
	\phi(p) = \arctan\Bigg[\frac{\hat{I}_1(p)\cos(\delta)-\hat{I}_2(p)} {\hat{I}_1(p)\sin(\delta)}\Bigg].
\end{equation}

%-------------------------------------------------------------------------------------------------
\subsection{Robust SLEF}
\label{ssec:SLEF-RE}
%-------------------------------------------------------------------------------------------------
\noindent In the previous subsection we estimate the parameters $\theta_1$ and $\theta_2$ with the LS method. This corresponds to assume a Gaussian distribution for the residual $\varepsilon$. By the examination of Figure \ref{fig:filt_gfb}, we noted that such residual is, in fact, non-Gaussian. Therefore, here we propose a robust procedure to improve the estimation of $\theta_1$ and $\theta_2$. Such robust estimator relies on the fact that the residual distribution has heavy tails \cite{rivera2016two, huber1992robust}. In general, the robust procedure can be formulated as the optimization problem:

\begin{equation}\label{eq:re1}
	\underset{\theta_1,\theta_2}{\arg\min}\sum_{p \in \mathcal{L}} \rho(\theta_1 x_i(p)^2 + \theta_2 y_i(p)^2 - 1; \kappa)
\end{equation}
where $\rho$ is a robust potential and $\kappa$ is a positive parameter that controls the outlier rejection sensitivity. In this paper we use the Leclerc's potential \cite{black1996unification}:

\begin{equation}\label{eq:rho}
	\rho(z;\kappa) = 1-\frac{1}{\kappa}\exp(-\kappa z^2)
\end{equation}
and we set $\kappa = 0.1$. 

According to \cite{huber1992robust, charbonnier1997deterministic, black1996unification}, the optimization can be obtained by the iteration of the solution of a weighted linear system; \emph{i.e.}, 

\begin{equation}\label{eq:T}
	T = (XWX^T)^{-1} XW\mathbbm{1}
\end{equation}
where $W$ is the diagonal matrix of weights

\begin{equation}\label{eq:W}
	W = \diag[w_1(x_1,y_1), w_2(x_2,y_2), \ldots, w_N(x_N,y_N)]
\end{equation}
with

\begin{equation}\label{eq:w}
	w_i(x_i,y_i) = \exp(-2k[\theta_1 x_i^2 + \theta_2 y_i^2 - 1]^2).
\end{equation} 

The solution $T^\ast$ is obtained by iterating \eqref{eq:T} and \eqref{eq:w}. The initial conditions for the weight matrix is set  $W = \diag[\mathbbm{1}]$. We observe that the system converged after just $3$ iterations.
% %-------------------------------------------------------------------------------------------------
\section{Experiments and results}
\label{sec:exps}
%-------------------------------------------------------------------------------------------------

\begin{figure}[ht]
	\centering
	
	\begin{subfigure}[ht]{0.19\linewidth}
		\centering
		\includegraphics[width=1\linewidth]{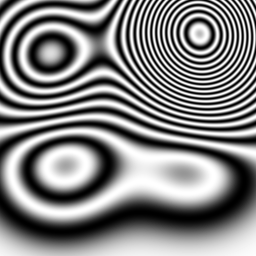}
		\caption{\centering}
		\label{fig:idini}
	\end{subfigure}
	\begin{subfigure}[ht]{0.19\linewidth}
		\centering
		\includegraphics[width=1\linewidth]{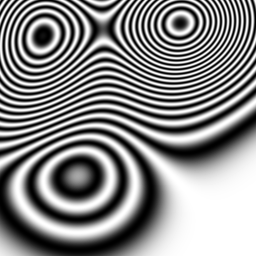}
		\caption{\centering}
	\end{subfigure}
	\begin{subfigure}[ht]{0.19\linewidth}
		\centering
		\includegraphics[width=1\linewidth]{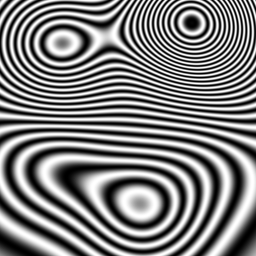}
		\caption{\centering}
	\end{subfigure}
	\begin{subfigure}[ht]{0.19\linewidth}
		\centering
		\includegraphics[width=1\linewidth]{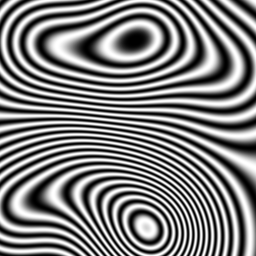}
		\caption{\centering}
	\end{subfigure}
	\begin{subfigure}[ht]{0.19\linewidth}
		\centering
		\includegraphics[width=1\linewidth]{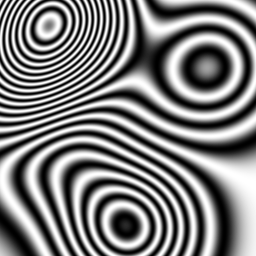}
		\caption{\centering}
	\end{subfigure}
	
	\begin{subfigure}[ht]{0.19\linewidth}
		\centering
		\includegraphics[width=1\linewidth]{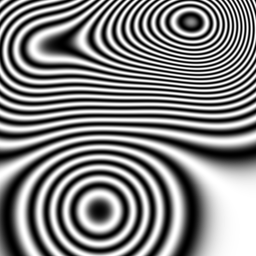}
		\caption{\centering}
	\end{subfigure}
	\begin{subfigure}[ht]{0.19\linewidth}
		\centering
		\includegraphics[width=1\linewidth]{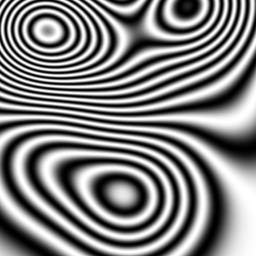}
		\caption{\centering}
	\end{subfigure}
	\begin{subfigure}[ht]{0.19\linewidth}
		\centering
		\includegraphics[width=1\linewidth]{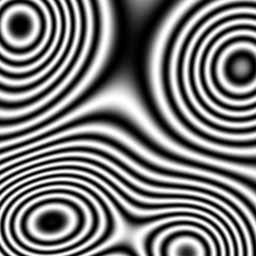}
		\caption{\centering}
	\end{subfigure}
	\begin{subfigure}[ht]{0.19\linewidth}
		\centering
		\includegraphics[width=1\linewidth]{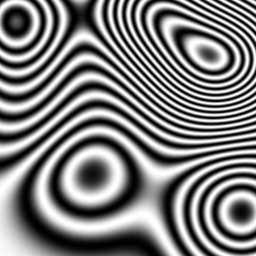}
		\caption{\centering}
	\end{subfigure}
	\begin{subfigure}[ht]{0.19\linewidth}
		\centering
		\includegraphics[width=1\linewidth]{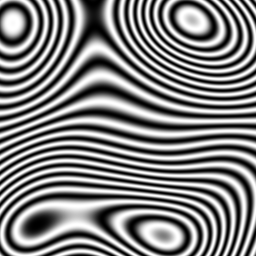}
		\caption{\centering}
		\label{fig:idfin}
	\end{subfigure}
	
	\begin{subfigure}[ht]{0.19\linewidth}
		\centering
		\includegraphics[width=1\linewidth]{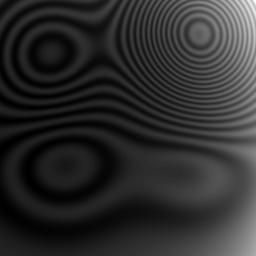}
		\caption{\centering $\sigma=0$}
		\label{fig:n01}
	\end{subfigure}
	\begin{subfigure}[ht]{0.19\linewidth}
		\centering
		\includegraphics[width=1\linewidth]{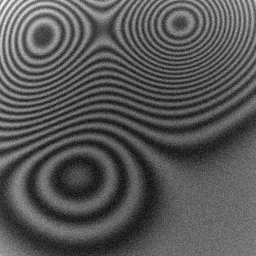}
		\caption{\centering $\sigma=0.25$}
		\label{fig:n11}
	\end{subfigure}
	\begin{subfigure}[ht]{0.19\linewidth}
		\centering
		\includegraphics[width=1\linewidth]{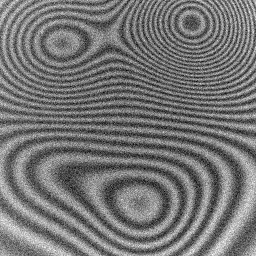}
		\caption{\centering  $\sigma=0.5$}
		\label{fig:n21}
	\end{subfigure}
	\begin{subfigure}[ht]{0.19\linewidth}
		\centering
		\includegraphics[width=1\linewidth]{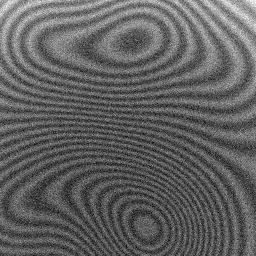}
		\caption{\centering  $\sigma=0.75$}
		\label{fig:n31}
	\end{subfigure}
	\begin{subfigure}[ht]{0.19\linewidth}
		\centering
		\includegraphics[width=1\linewidth]{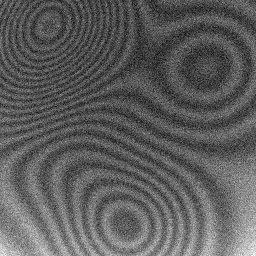}
		\caption{\centering  $\sigma=1.0$}
		\label{fig:n41}
	\end{subfigure}
	
	\begin{subfigure}[ht]{0.19\linewidth}
		\centering
		\includegraphics[width=1\linewidth]{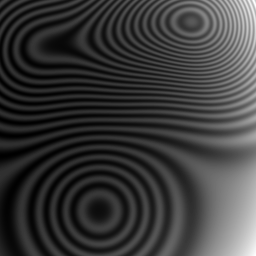}
		\caption{\centering  $\sigma=0$}
		\label{fig:n02}
	\end{subfigure}
	\begin{subfigure}[ht]{0.19\linewidth}
		\centering
		\includegraphics[width=1\linewidth]{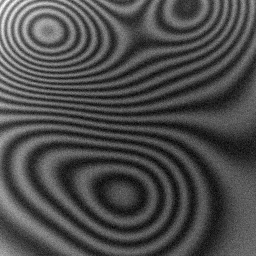}
		\caption{\centering  $\sigma=0.25$}
		\label{fig:n12}
	\end{subfigure}
	\begin{subfigure}[ht]{0.19\linewidth}
		\centering
		\includegraphics[width=1\linewidth]{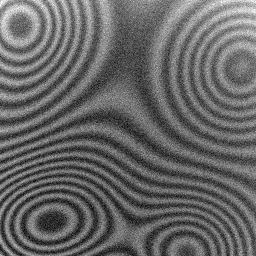}
		\caption{\centering  $\sigma=0.5$}
		\label{fig:n22}
	\end{subfigure}
	\begin{subfigure}[ht]{0.19\linewidth}
		\centering
		\includegraphics[width=1\linewidth]{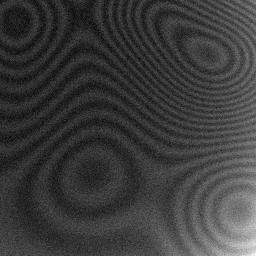}
		\caption{\centering  $\sigma=0.75$}
		\label{fig:n32}
	\end{subfigure}
	\begin{subfigure}[ht]{0.19\linewidth}
		\centering
		\includegraphics[width=1\linewidth]{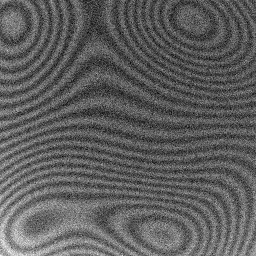}
		\caption{\centering  $\sigma=1.0$}
		\label{fig:n42}
	\end{subfigure}
	
	\caption{Synthetic patterns.}
	\label{fig:patterns}
\end{figure}

\noindent In order to evaluate the proposed algorithms performance, we use ten sets of synthetic patterns of $512 \times 512$ with different phase steps $(\delta = [\pi/10, \pi/6, \pi/4, \pi/3, \pi/2])$ and five different Gaussian noise levels $(\sigma = [0.0, 0.25, 0.5, 0.75, 1.0])$ . These patterns present spatial and temporal dependency of background intensities, amplitude modulations and noise. In Figure \ref{fig:patterns} we present the used patterns. Figures \ref{fig:idini} to \ref{fig:idfin} are the normalized noiseless patterns. It can be observed that the patterns high and low frequencies with circular fringes. Our main interest are these kind of patterns (circular ones) due to the sign error induced in the phase estimation, reason why we are using a two steps algorithm.

Figures \ref{fig:n01} to \ref{fig:n42} present the different noise levels applied to each synthetic fringe pattern as well as the background variations and the amplitude modulations. For illustrative purposes we present a different noise level applied to a different pattern, nevertheless, all the noise levels as well as the background variations and amplitude modulations were applied to all the samples. 

In order to compare our proposal, we performed the phase step estimation by pre--filtering the synthetic patterns with the HHT as proposed in \cite{liu2016simultaneous} (which we call LEF--HHT) by using the implementation the Enhaced Fast Empirical Decomposition (EFEMD) proposed in \cite {trusiak2014advanced, bhuiyan2008fast}. 

Figure \ref{fig:ideal} depicts the Lissajous Pattern (LP) for noiseless and normalized fringe patterns with a phase shift of $\delta = \pi/3$; in this case, the LP is centered at the origin and it is perfectly well--marked. In order to remove the rotation of the ellipse, as seen in Figure \ref{fig:lef}, the LP is computed by using the addition and the subtraction of the fringe patterns  \cite{farrell1992phase}. These ideal figures correspond to the pattern presented in Figure \ref{fig:idfin}. In Figure \ref{fig:var} we present the same pattern with spatial and temporal dependency of the background and amplitude modulation as well as Gaussian noise of $\sigma = 0.5$, it is clear that a LP is not appreciated from the original data given the disturbances previously mentioned.

%-------------------------------------------------------------------------------------------------
\subsection{Pre--filtering process}
\label{ssec:prefilt}
%-------------------------------------------------------------------------------------------------

\begin{figure}[ht]
	\centering
	\begin{subfigure}[ht]{\linewidth}
		\centering
		\includegraphics[width=.32\linewidth]{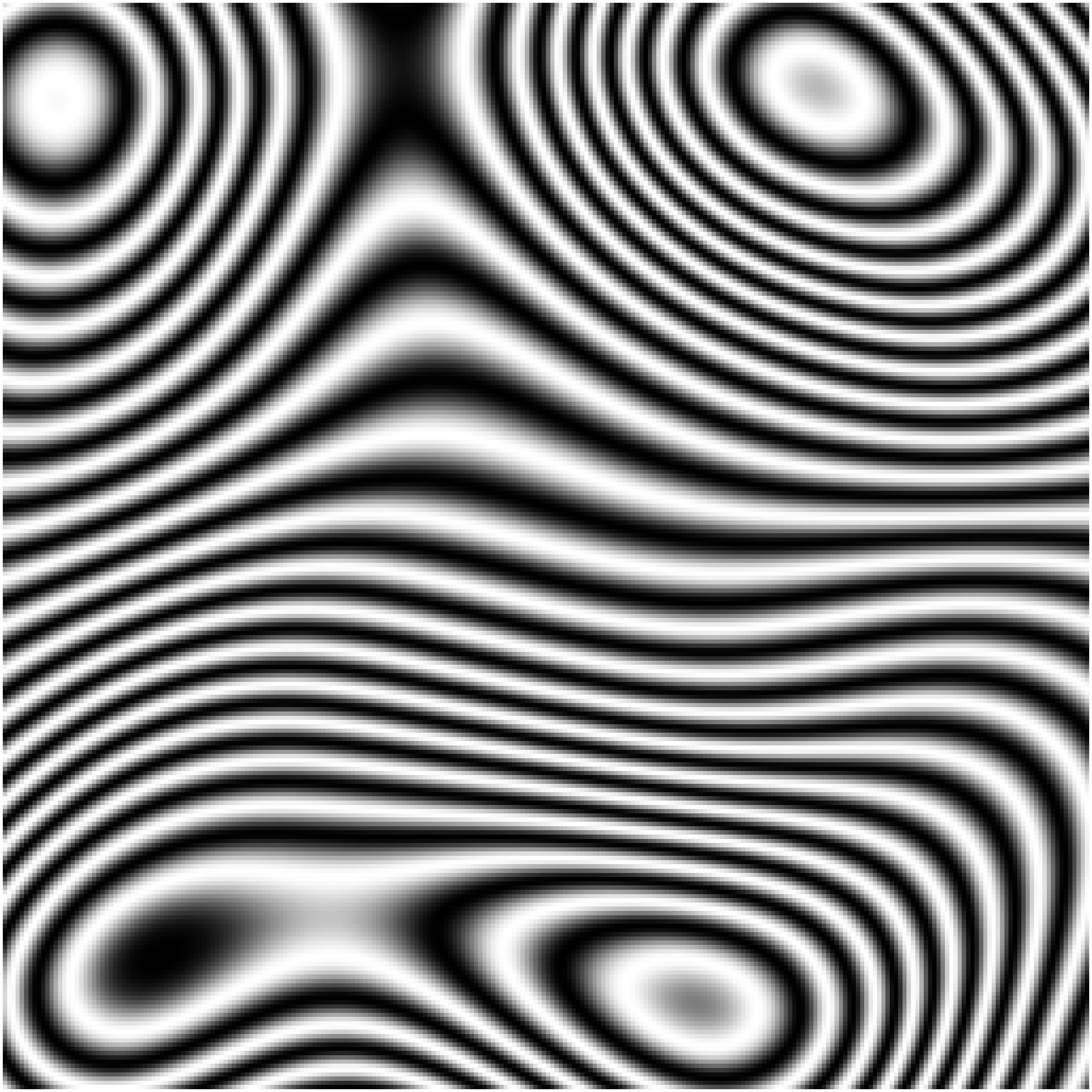}
		\includegraphics[width=.32\linewidth]{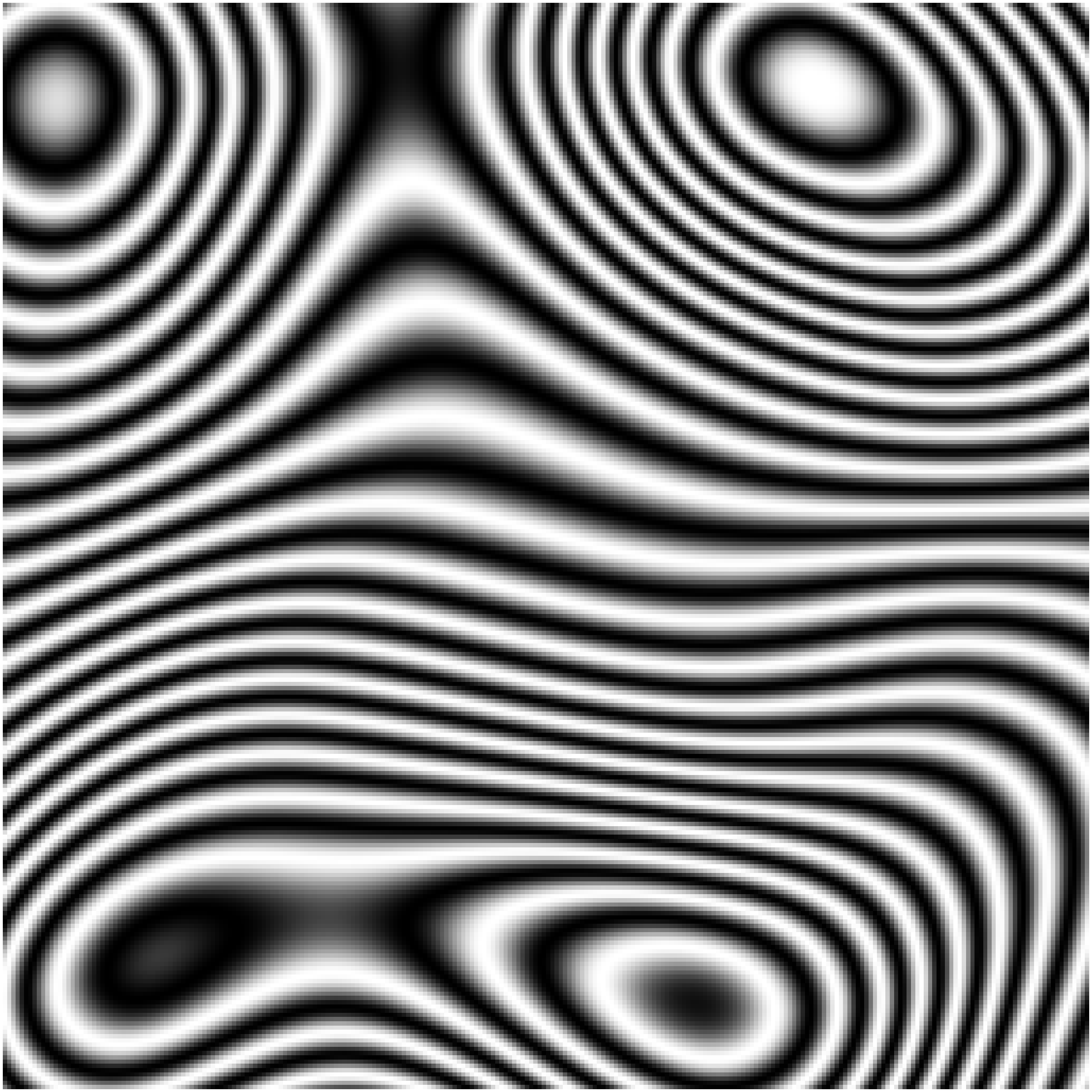}
		\includegraphics[width=.32\linewidth]{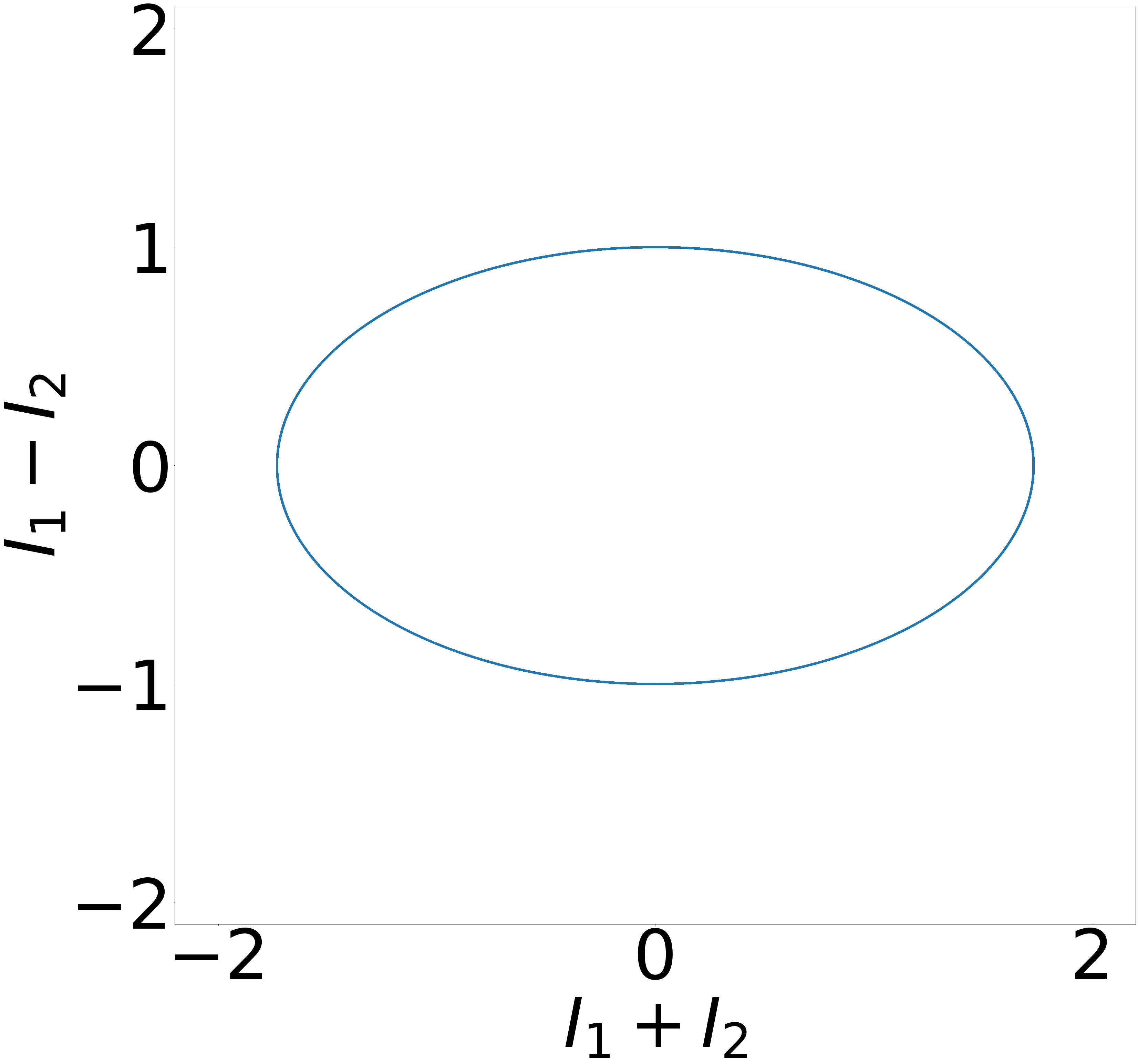}
		\caption{\centering Ideal patterns}
		\label{fig:ideal}
	\end{subfigure}
	
	\begin{subfigure}[ht]{\linewidth}
		\centering
		\includegraphics[width=.32\linewidth]{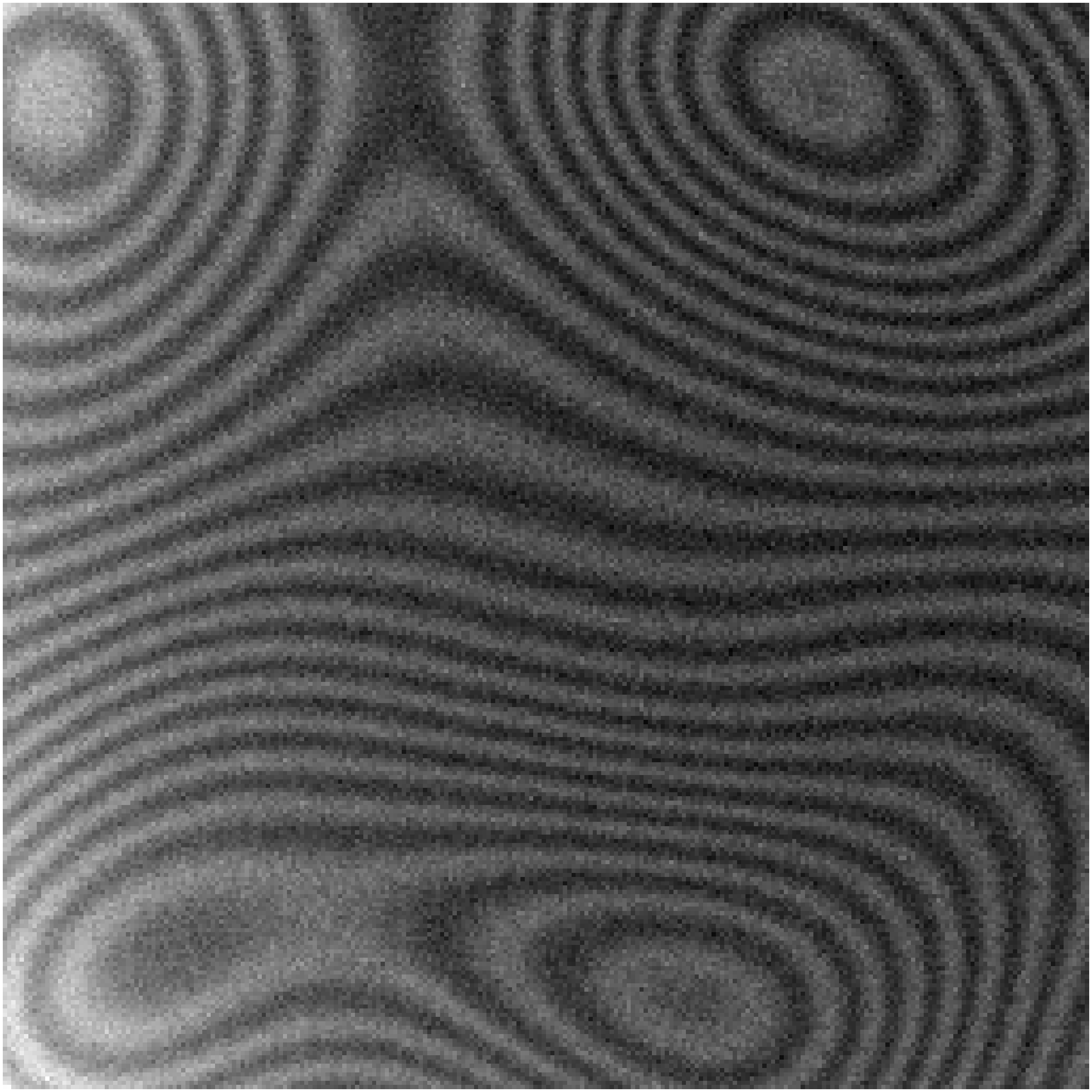}
		\includegraphics[width=.32\linewidth]{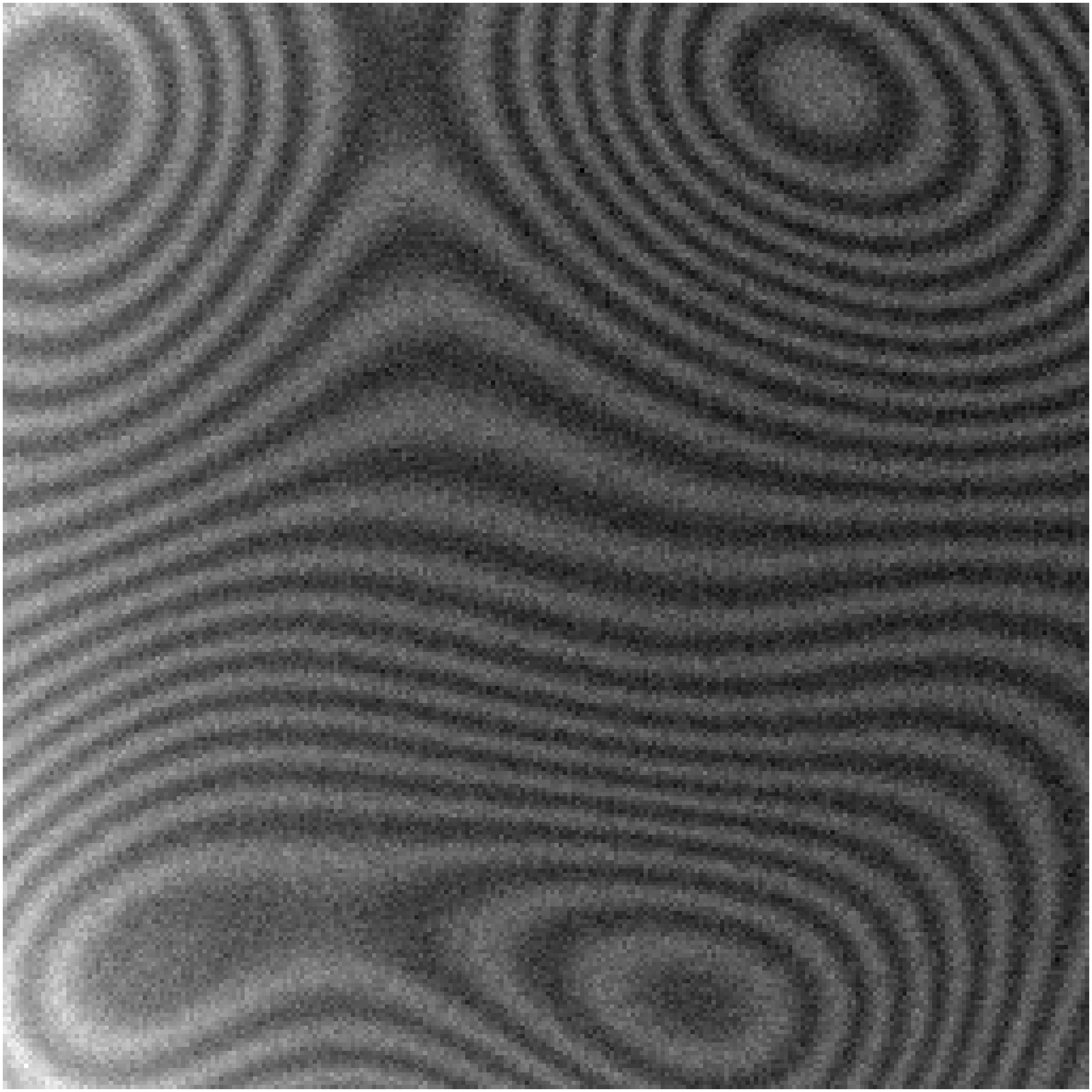}
		\includegraphics[width=.32\linewidth]{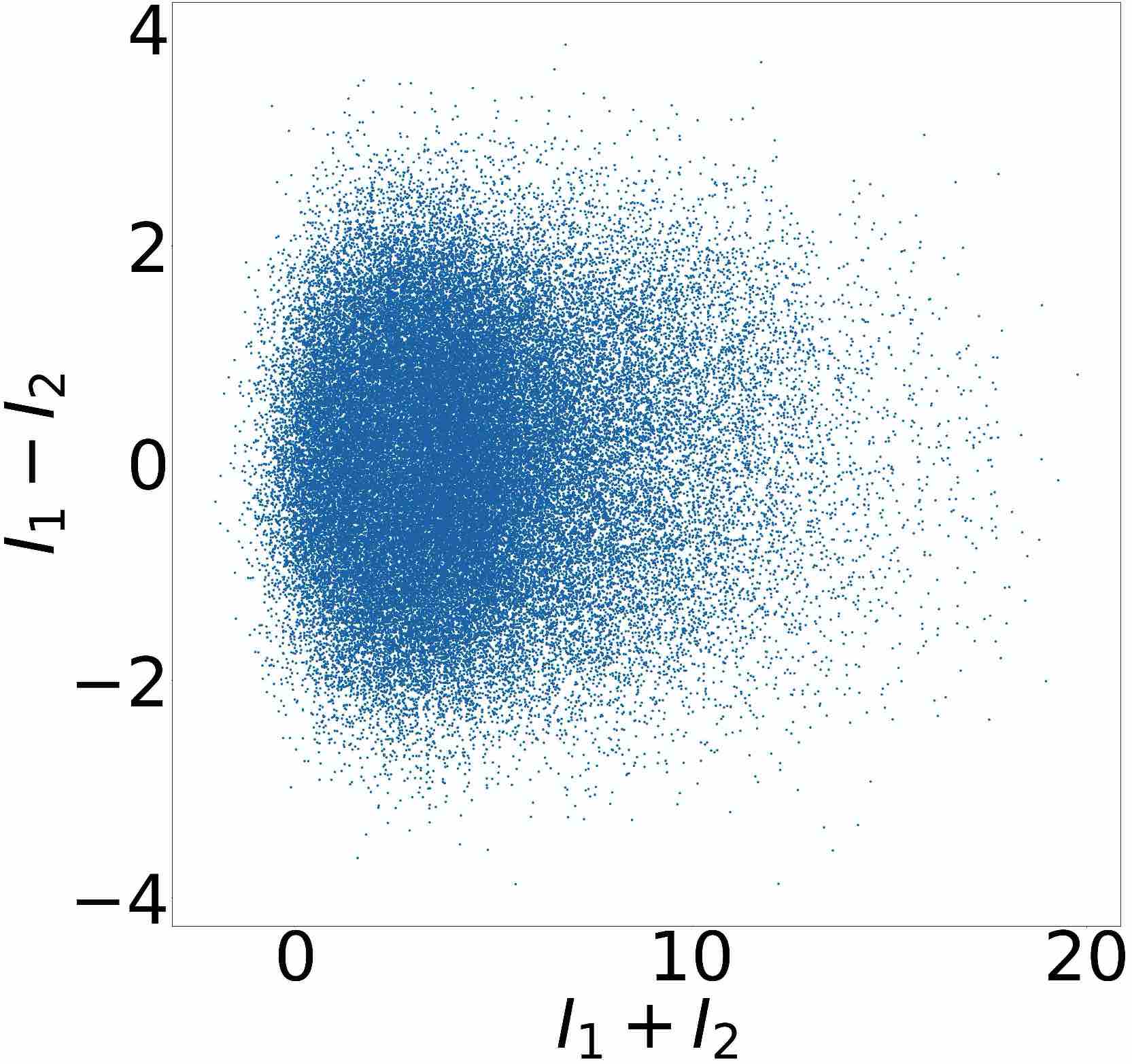}
		\caption{\centering Experimental patterns}
		\label{fig:var}
	\end{subfigure}
	
	\begin{subfigure}[ht]{\linewidth}
		\centering
		\includegraphics[width=.32\linewidth]{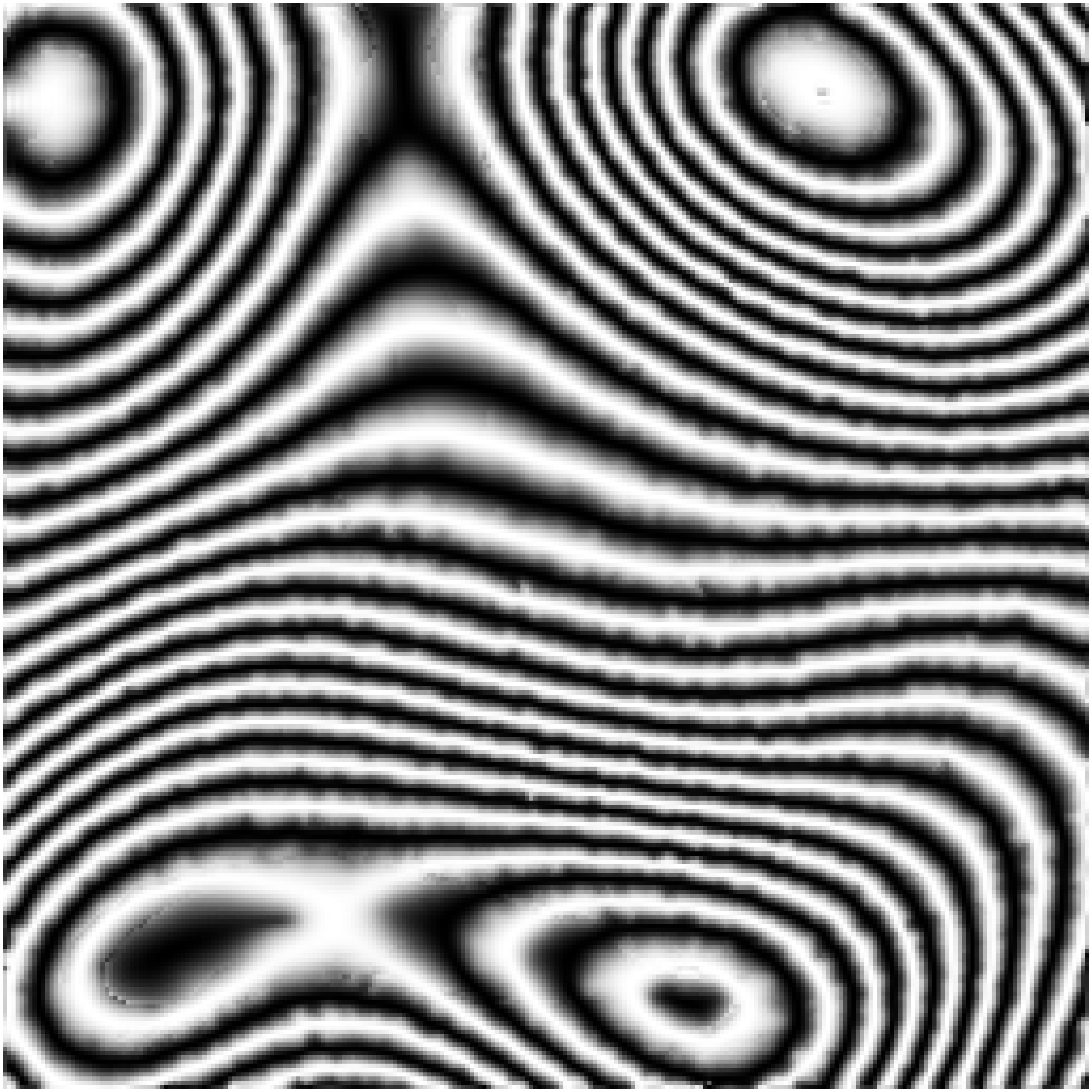}
		\includegraphics[width=.32\linewidth]{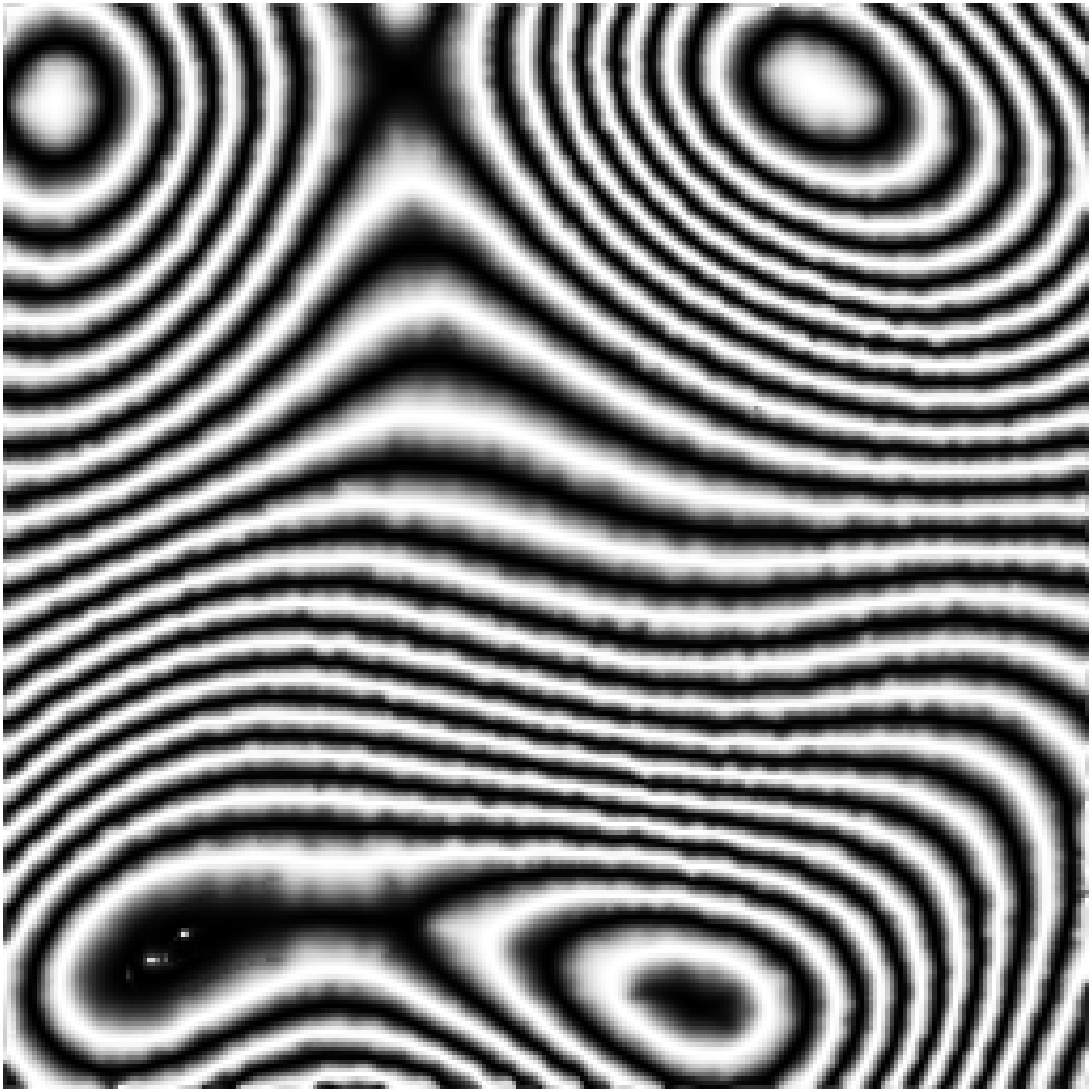}
		\includegraphics[width=.32\linewidth]{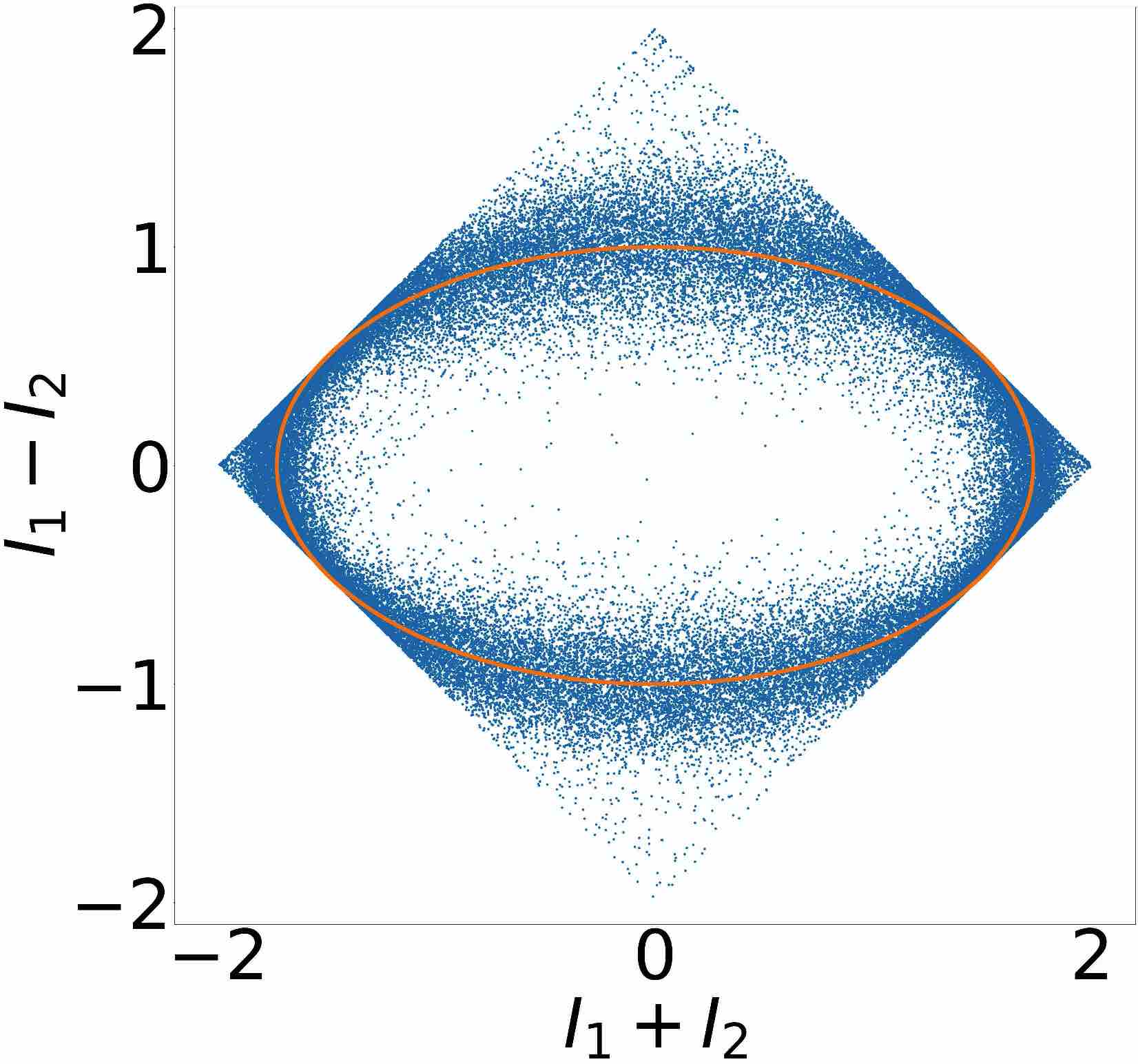}
		\caption{\centering Filtered patterns using HHT}
		\label{fig:filt_hht}
	\end{subfigure}
	
	\begin{subfigure}[ht]{\linewidth}
		\centering
		\includegraphics[width=.32\linewidth]{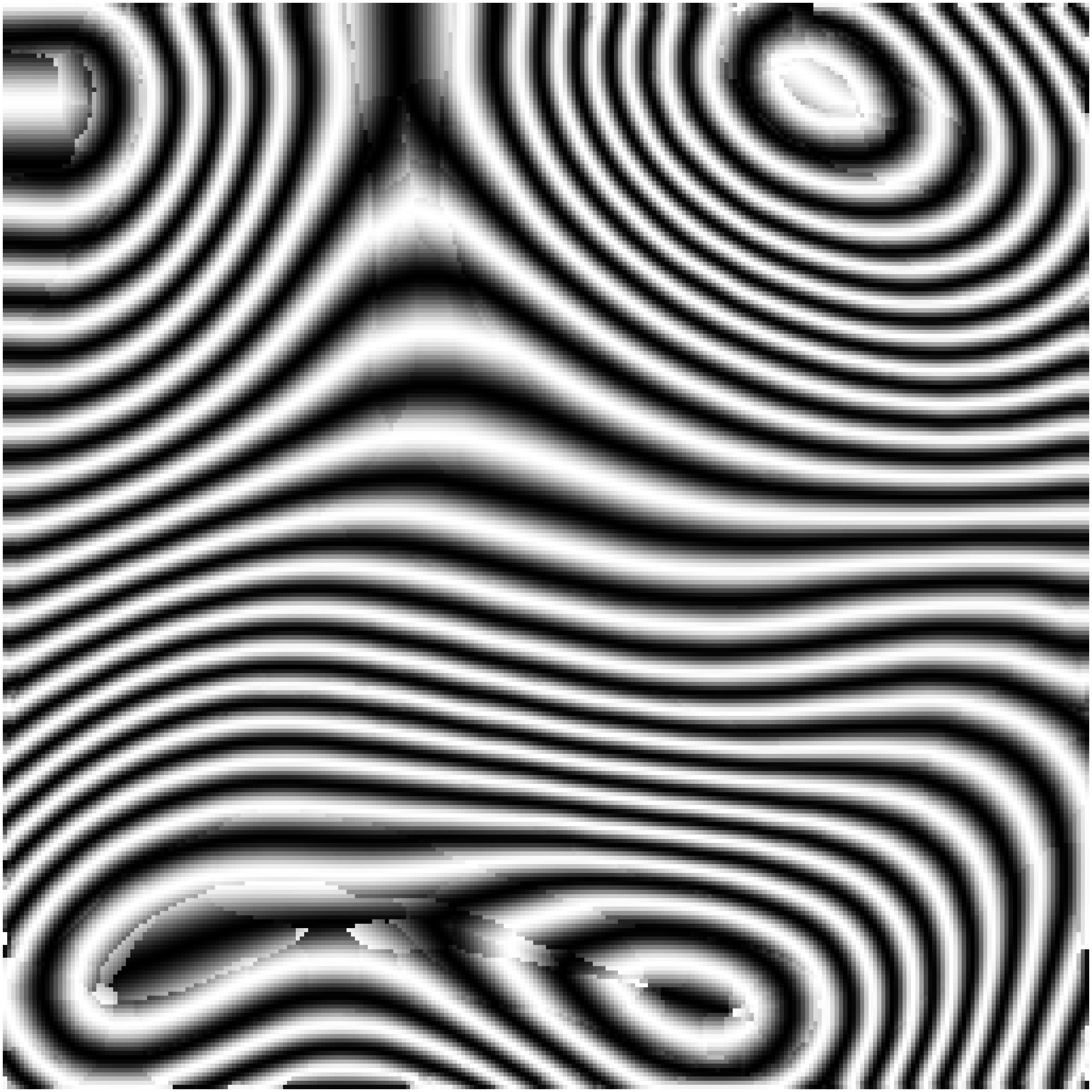}
		\includegraphics[width=.32\linewidth]{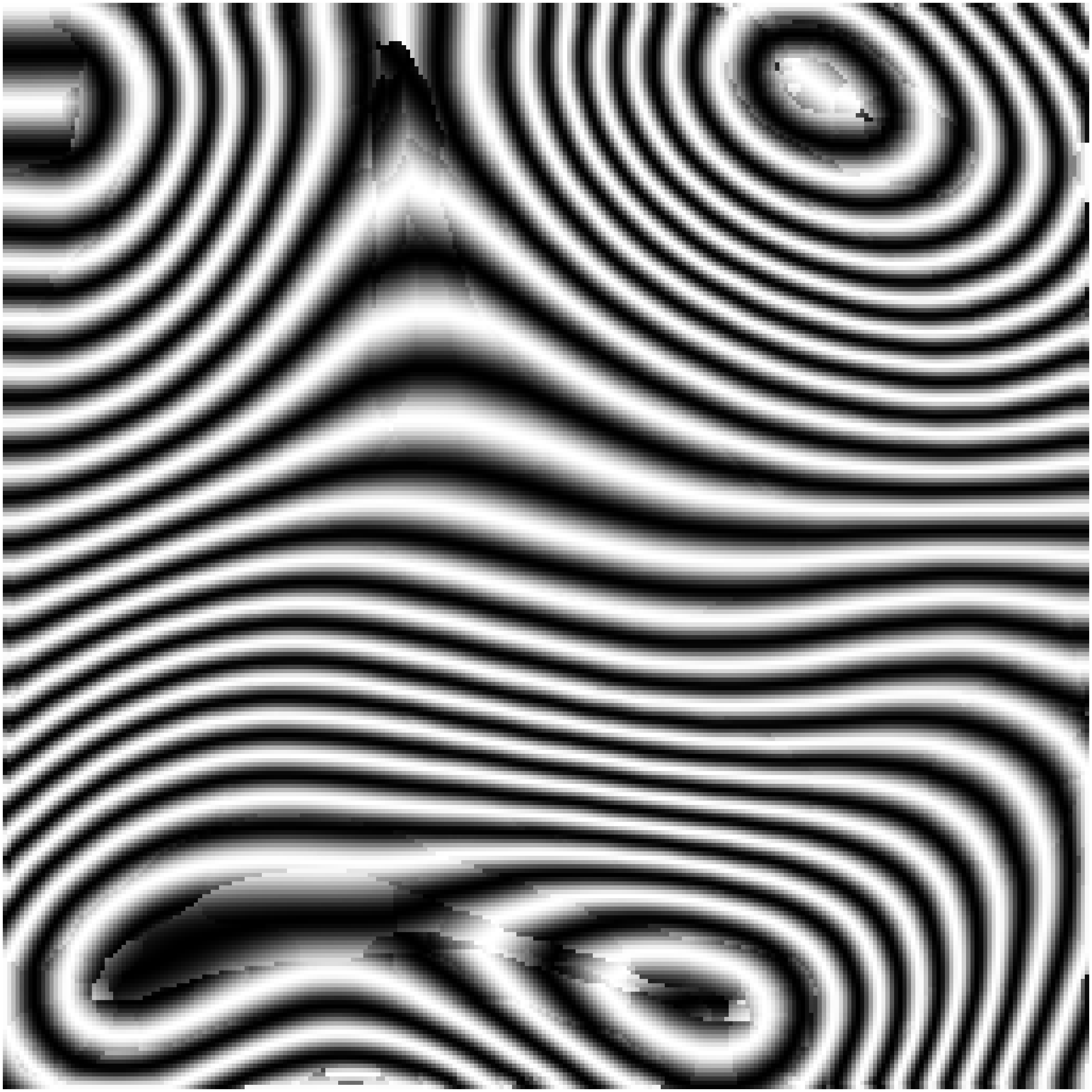}
		\includegraphics[width=.32\linewidth]{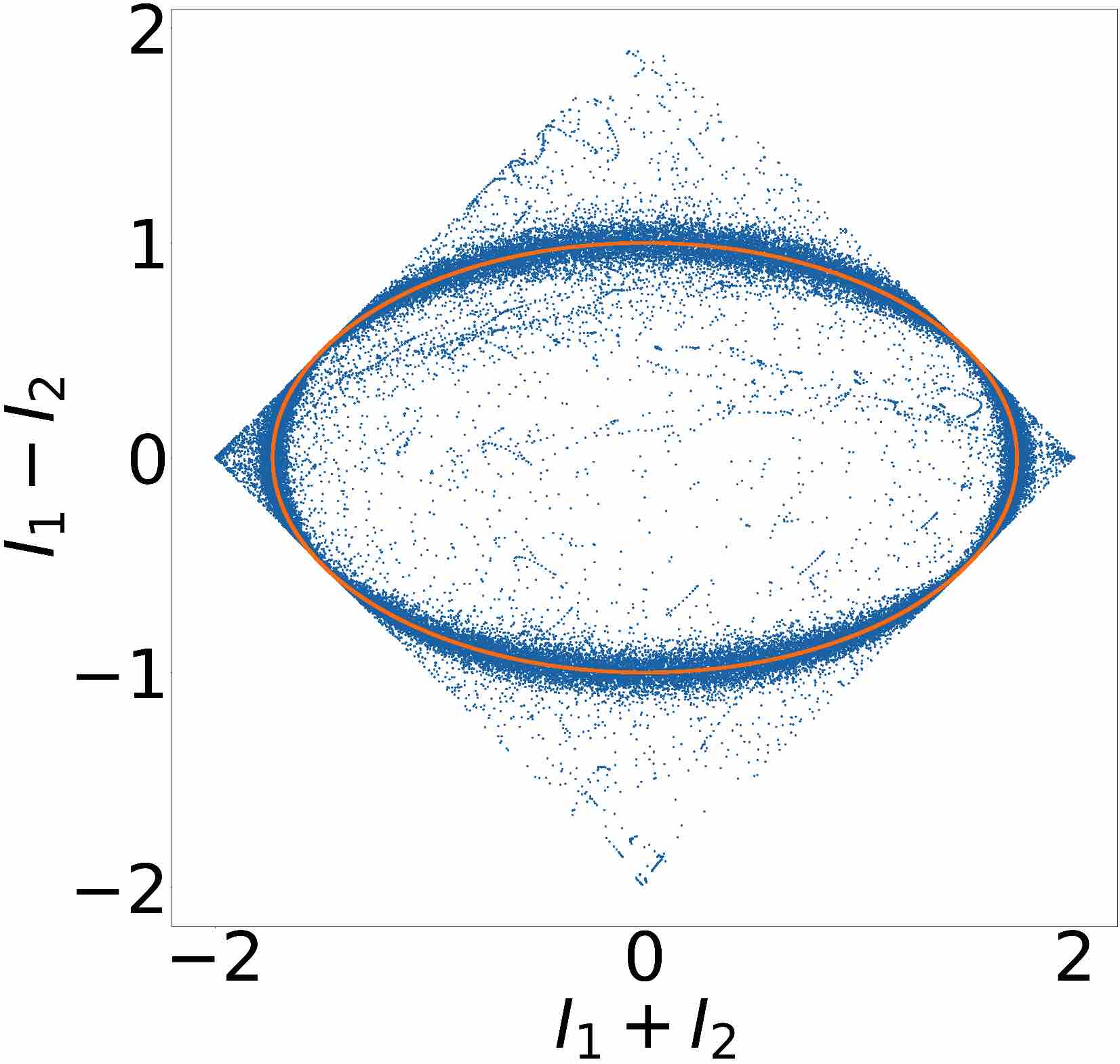}
		\caption{\centering Filtered patterns using GFB}
		\label{fig:filt_gfb}
	\end{subfigure}
	
	\caption{Lissajous figures for different interferogram cases. The phase step between the patterns is $\delta = \pi/3$.}
	\label{fig:lef2}
\end{figure}

The first step of our method consists on the preprocessing of the fringe patterns. For comparison purposes, in Figure \ref{fig:filt_hht} we show the filtered patterns using the HHT with its respective LP. It can be seen that the background term is retrieved but some level of noise remains. In this example, we applied a decomposition of seven modes with a sifting equal to $.01$. On the other hand, the LP pattern presents a dispersed cloud of points, nevertheless, the shape and the eccentricity of the ellipse is still noticeable. Finally, in Figure \ref{fig:filt_gfb} we present the filtered pattern using the GFB and its respective LP. Even though the noise filtering is better as well as the normalization, there are several residuals as stablished in \eqref{eq:eli4n}.

From obtained LP of the GFB filtered patterns, it is important to emphasize the following:

\begin{enumerate}
	\item The LP center is at the origin.
	\item The approximation of the points is close to the ideal ellipse.
	\item The spread points are due to the residuals of the preprocess.
\end{enumerate}

%-------------------------------------------------------------------------------------------------
\subsection{Phase step calculation}
\label{ssec:step}
%-------------------------------------------------------------------------------------------------
The second step is the estimation of the phase step $\delta$ using \eqref{eq:delta}. To evaluate the feasibility of our proposed methods, we calculated the phase step estimation comparing them with the preprocessed LEF algorithm proposed in \cite{liu2016simultaneous}. The first comparison consists on calculating the phase step at different Gaussian noise levels. For this analysis, we used ten different sets of images (See Figure \ref{fig:patterns}) with five different Gaussian noise levels $(\sigma = [0.0, 0.25, 0.5, 0.75, 1.0])$, with spatial--temporal variations in their background intensity as well as the amplitude modulation. The phase step between the patterns was set to $\pi/3$. Each pattern was preprocessed using the HHT and the calculation of the phase step was using the same \eqref{eq:delta} using the computed parameters of each method: LEF--HHT, SLEF--LS and SLEF--RE .

\begin{figure}[ht]
	\centering
	\includegraphics[width=1\linewidth]{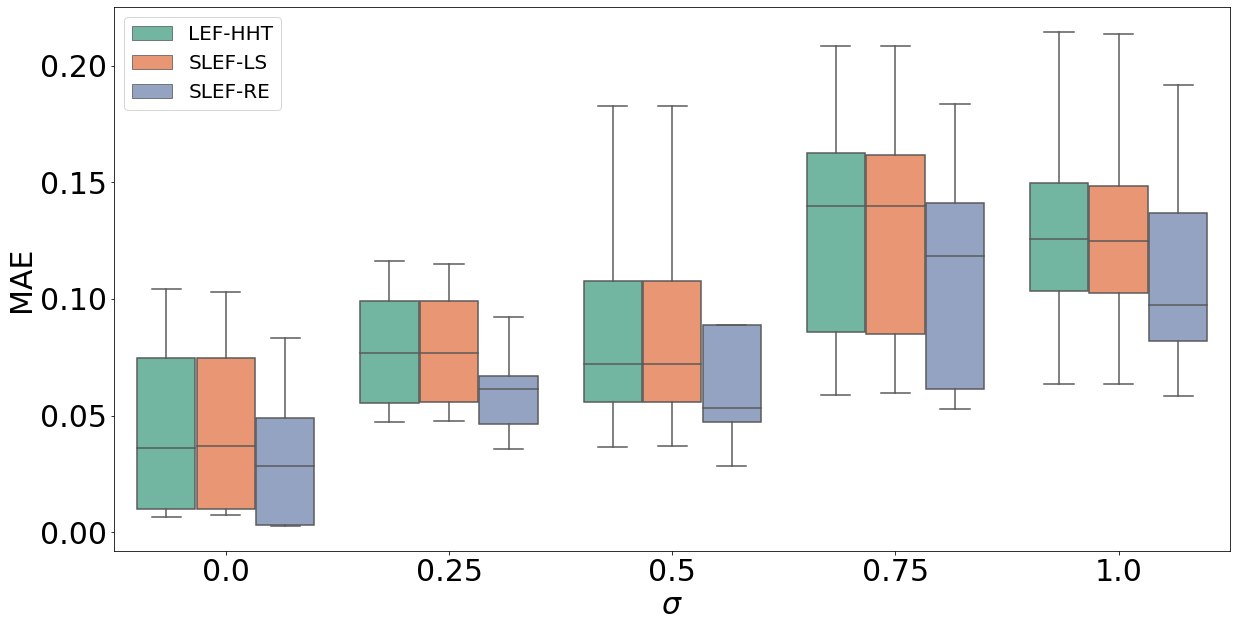}
	\caption{Mean Absolute Error of the phase step calculation using HHT preprocess for the LEF--HHT, SLEF--LS and SLEF--RE algorithms at different noise levels.}
	\label{fig:mae_hht}
\end{figure}

Figure \ref{fig:mae_hht} depicts the computed Mean Absolute Error (MAE) of the analyzed sets of patterns pre--filtered with the HHT method. The total mean error of the algorithms are: $MAE_{LEF-HHT} = 0.1021 rad$, $MAE_{SLEF-LS} = 0.1018 rad$ and $MAE_{SLEF-RE} =0.0991 rad$. It can be appreciated that the algorithms present similar behavior at high noise levels while the SLEF algorithms reduce the variance at low noise levels. The MAE in noise levels superior to $\sigma=0.5$ is around $0.1 rad$ which could generate the presence of harmonics in the recovered phase.

Now, we pre--filtered the same synthetic patterns with the GFB to demonstrate the feasibility of this technique. The tested algorithms are the 5--term LEF which we will call LEF--GFB and the prosed SLEF--LS and SLEF--RE.

\begin{figure}[ht]
	\centering
	\includegraphics[width=1\linewidth]{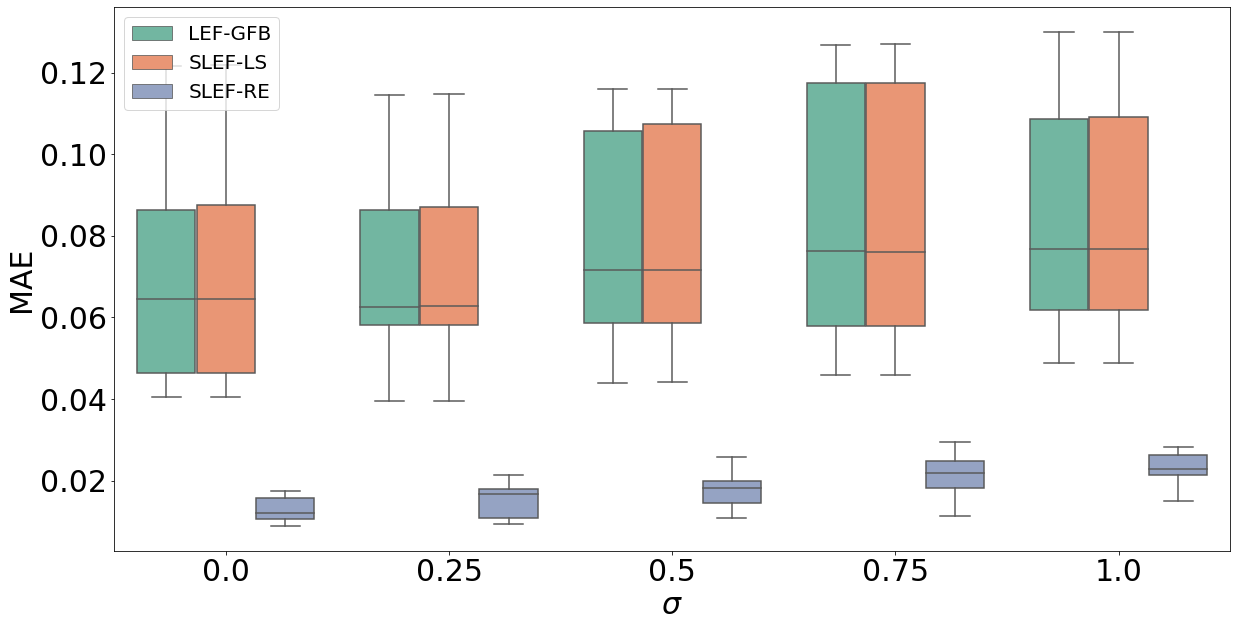}
	\caption{Mean Absolute Error of the phase step calculation using GFB preprocess for the LEF--GFB, SLEF--LS and SLEF--RE algorithms at different noise levels.}
	\label{fig:mae_gfb}
\end{figure}

In Figure \ref{fig:mae_gfb} we present the MAE of the analyzed sets of pre--filtered patterns. The total mean error of the algorithms are: $MAE_{LEF-GFB} = 0.0869 rad$, $MAE_{SLEF-LS} = 0.0871rad$ and $MAE_{SLEF-RE} =0.0204 rad$. It can be appreciated that the well-known LEF algorithm and the proposed SLEF--LS algorithm have practically the same behavior, meaning that if the patterns are well normalized, these algorithms are equivalent; with the SLEF--LS's advantage of having simplified solution. Moreover, SLEF--RE is a more accurate (less error) and precise (less error variance) phase step estimator. Regardless of the noise presented, the MAE is smaller than $0.025 rad$ in all levels.

The results presented in Figure \ref{fig:mae_gfb} prove that the GFB pre--filtering is more robust to high noise levels, which improve the stability of the algorithms. On lower noise levels the performance is similar on both pre--filtering process. The SLEF--RE algorithm has the best performance when using GFB as normalizing method.

To evaluate the estimation of our method, we also performed a test consisting on the calculation of the phase step with different phase steps $\delta$. For this analysis, we used the same images with a fixed Gaussian noise of $\sigma = 0.5$. The phase steps between the patterns were $\delta = [\pi/10, \pi/6, \pi/4, \pi/3, \pi/2]$. As before, each pattern was preprocessed using the GFB and the calculation of the phase step was using the same  \eqref{eq:delta} with the computed parameters of each method: SLEF--LS, SLEF--RE and LEF--GFB.

\begin{figure}[ht]
	\centering
	\includegraphics[width=1\linewidth]{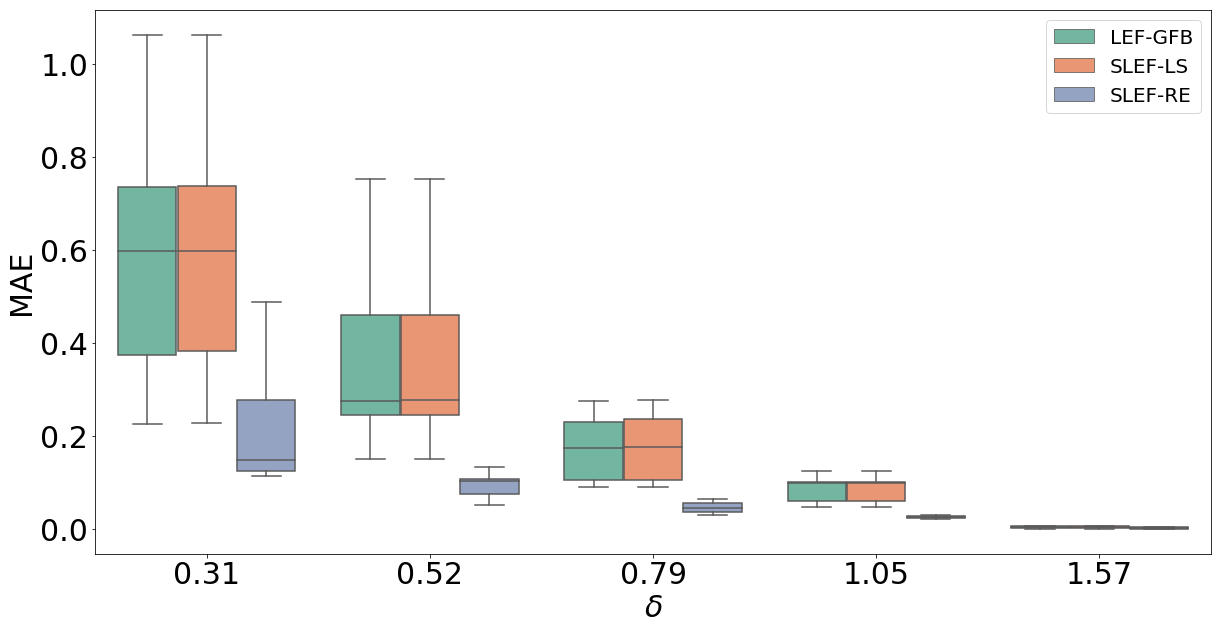}
	\caption{Mean Absolute Error of the phase step calculation for the LEF--GFB SLEF--LS and SLEF--RE algorithms at different phase steps.}
	\label{fig:shiftmae}
\end{figure}

Figure \ref{fig:shiftmae} shows the MAE resulted of the analyzed patterns. It can be observed that the LEF--GFB and SLEF--LS algorithms present the same behavior, proving their equivalency, while SLEF--RE have smaller variance and error. It is important to notice that the error increases for small phase steps, and it decreases as the step approaches to $\delta=\pi/2$. The error for phase steps in the interval $\delta \in (\pi/2,\pi)$ is the same as the presented in Figure \ref{fig:shiftmae}, since the error tends to increase as the step gets closer to $\pi$.

\begin{figure}[ht]
	\centering
	\includegraphics[width=1\linewidth]{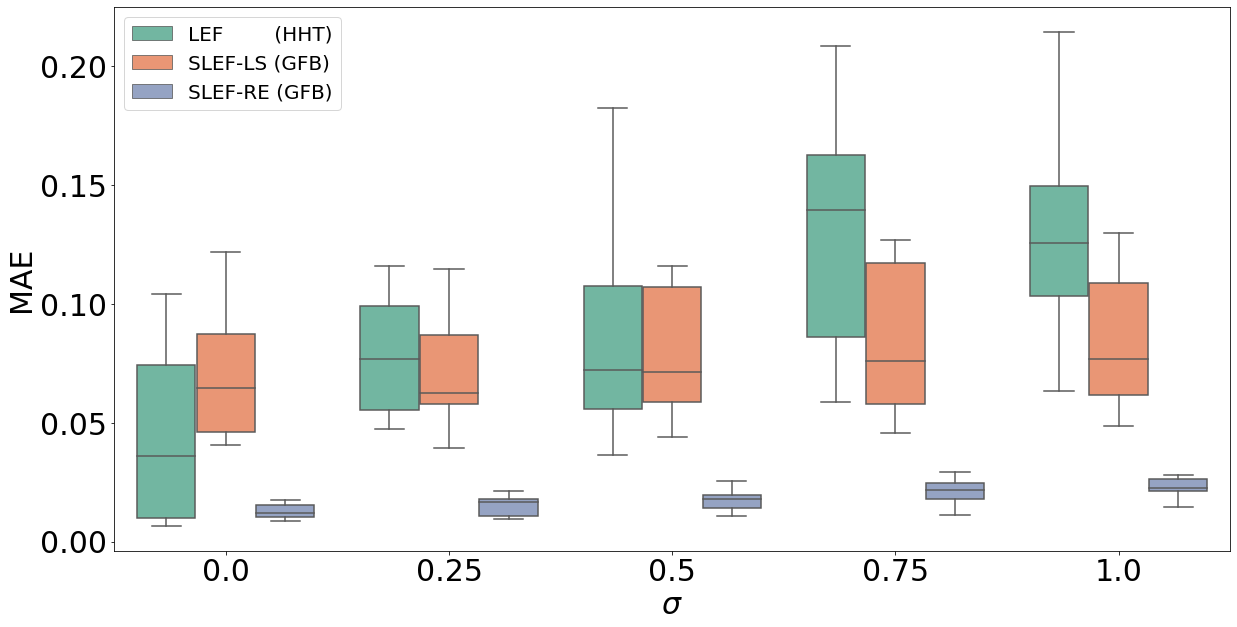}
	\caption{Comparison of the 5--term LEF algorithm with HHT pre--filtering with the proposed algorithms SLEF--LS and SLEF--RE.}
	\label{fig:gfb_hht}
\end{figure}

We include Figure \ref{fig:gfb_hht} to demonstrate the behavior of the LEF--HHT algorithm and our GFB based proposals. As mentioned before, the LEF--HHT performs better than the SLEF--LS in low noise level. The SLEF--RE demonstrated to be de most accurate the most accurate in all noise levels.

%-------------------------------------------------------------------------------------------------
\subsection{Phase map estimation}
\label{ssec:phase}
%-------------------------------------------------------------------------------------------------

Finally, the third stage is the calculation of the phase using Eq. \eqref{eq:phi1} for the LEF algorithm and Eq. \eqref{eq:phi2} for the SLEF algorithms. To evaluate the phase error, we used the pattern presented in Figure \ref{fig:lef2}. In this case, the phase shift is $\delta = \pi/3$ and the Gaussian noise presents a $\sigma = 0.5$. The purpose of this experiment is to prove that the piston term induced in \eqref{eq:phi1} can be avoided by using \eqref{eq:phi2} instead. The fringe patterns were normalized with GFBs and since we already demonstrated that SLEF--LS and LEF are equivalent given such condition, we assume that the  phase estimation using \eqref{eq:phi2} would result the same with 2--terms or 5--terms.

\begin{figure*}[ht]
	\centering
	\begin{subfigure}[ht]{0.19\linewidth}
		\centering
		\includegraphics[width=1\linewidth]{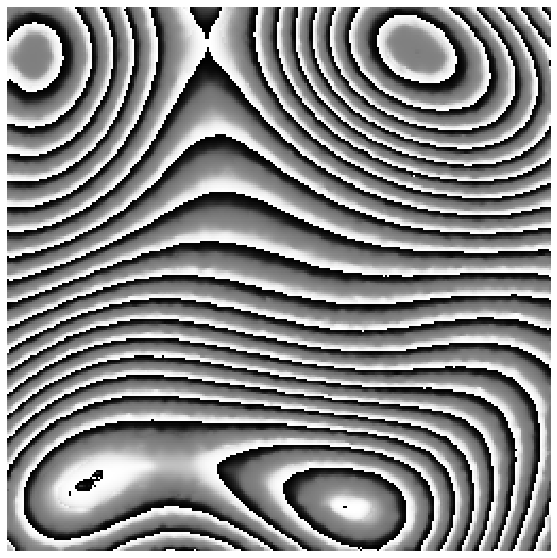}
		\caption{\centering Phase LEF-HHT}
		\label{fig:phi_lefhht}
	\end{subfigure}
	\begin{subfigure}[ht]{0.19\linewidth}
		\centering
		\includegraphics[width=1\linewidth]{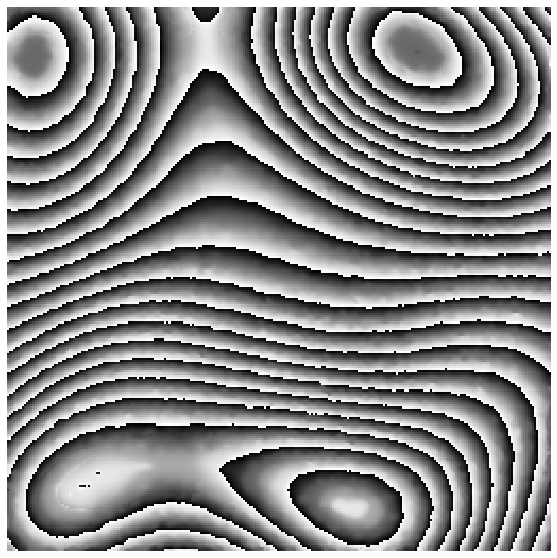}
		\caption{\centering Phase LEF-HHT-2}
		\label{fig:phi_lefhht_mur}
	\end{subfigure}
	\begin{subfigure}[ht]{0.19\textwidth}
		\centering
		\includegraphics[width=1\linewidth]{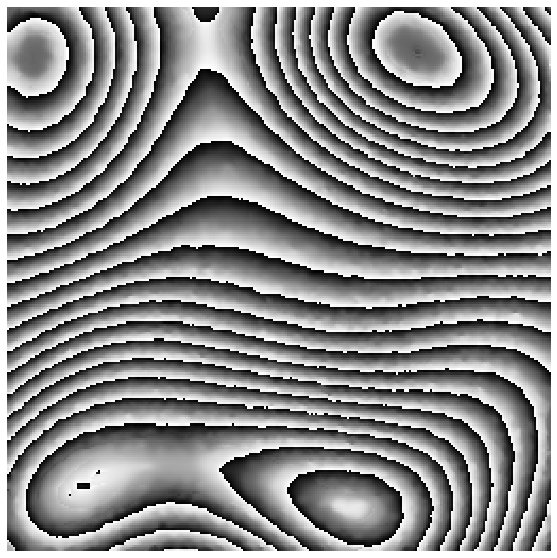}
		\caption{\centering Phase SLEF-HHT}
		\label{fig:phi_lefhh}
	\end{subfigure}
	\begin{subfigure}[ht]{0.19\linewidth}
		\centering
		\includegraphics[width=1\linewidth]{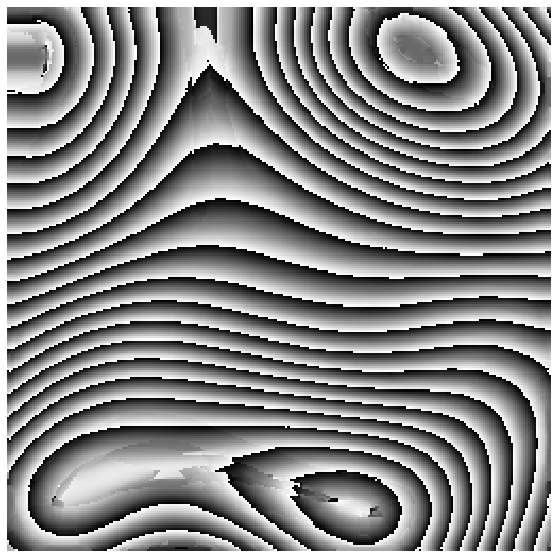}
		\caption{\centering Phase SLEF-LS}
		\label{fig:phi_slefls}
	\end{subfigure}
	\begin{subfigure}[ht]{0.19\linewidth}
		\centering
		\includegraphics[width=1\linewidth]{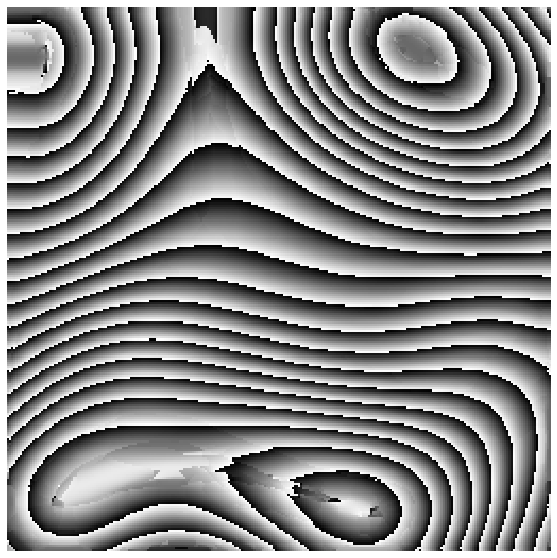}
		\caption{\centering Phase SLEF-RE}
		\label{fig:phi_slefre}
	\end{subfigure}
	
	\begin{subfigure}[ht]{0.19\linewidth}
		\centering
		\includegraphics[width=1\linewidth]{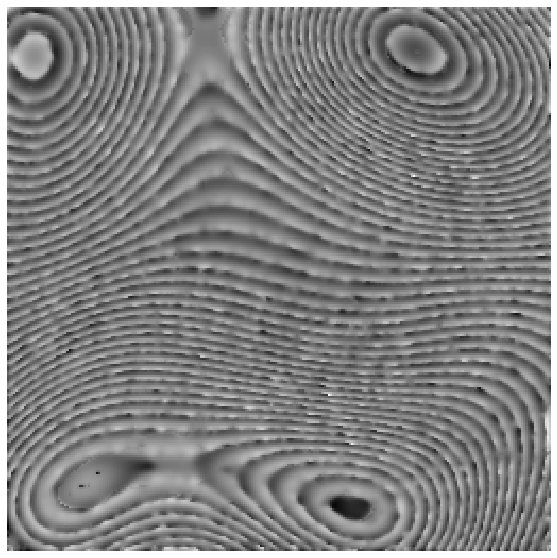}
		\caption{\centering Error LEF-HHT}
		\label{fig:error_lefhht}
	\end{subfigure}
	\begin{subfigure}[ht]{0.19\linewidth}
		\centering
		\includegraphics[width=1\linewidth]{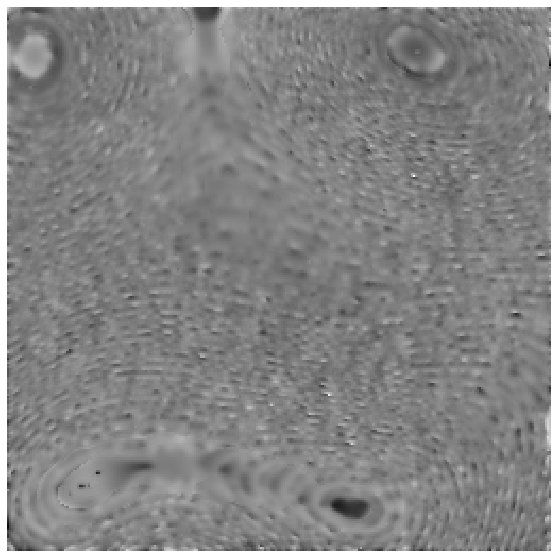}
		\caption{\centering Error LEF-HHT-2}
		\label{fig:error_lefhht}
	\end{subfigure}
	\begin{subfigure}[ht]{0.19\linewidth}
		\centering
		\includegraphics[width=1\linewidth]{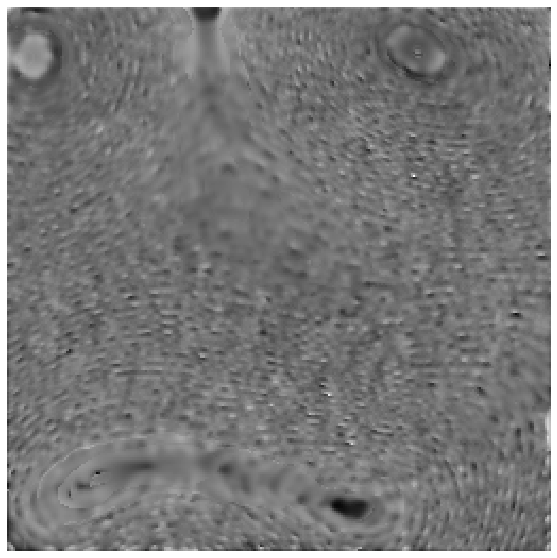}
		\caption{\centering Error SLEF-HHT}
		\label{fig:error_lefhh}
	\end{subfigure}
	\begin{subfigure}[ht]{0.19\linewidth}
		\centering
		\includegraphics[width=1\linewidth]{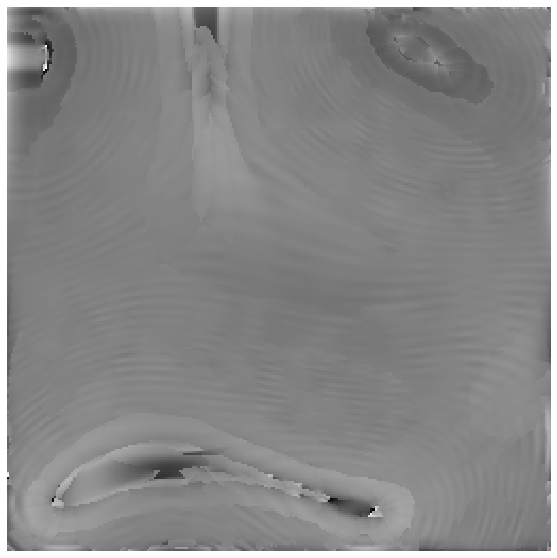}
		\caption{\centering Error SLEF-LS}
		\label{fig:error_slefls}
	\end{subfigure}
	\begin{subfigure}[ht]{0.19\linewidth}
		\centering
		\includegraphics[width=1\linewidth]{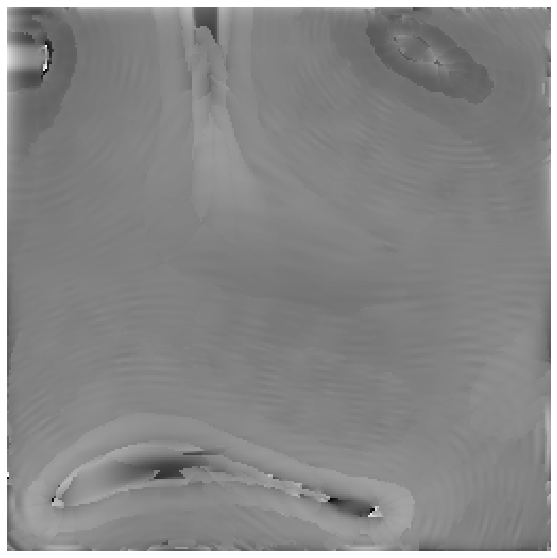}
		\caption{\centering Error SLEF-RE}
		\label{fig:error_slefre}
	\end{subfigure}
	
	\begin{subfigure}[ht]{0.19\linewidth}
		\centering
		\includegraphics[width=1\linewidth]{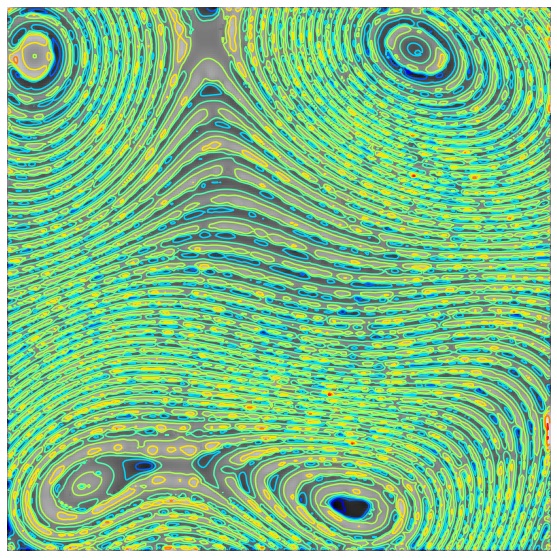}
		\caption{\centering Error LEF-HHT (contour)}
		\label{fig:error_lefhhtc}
	\end{subfigure}
	\begin{subfigure}[ht]{0.19\linewidth}
		\centering
		\includegraphics[width=1\linewidth]{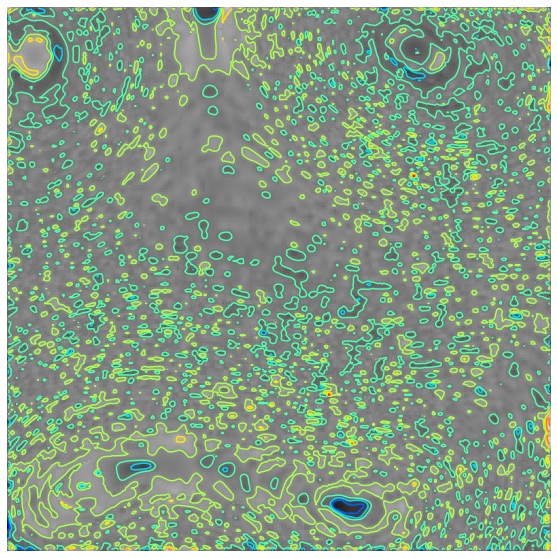}
		\caption{\centering Error LEF-HHT-2 (contour)}
		\label{fig:error_lefhhtc_mur}
	\end{subfigure}
	\begin{subfigure}[ht]{0.19\linewidth}
		\centering
		\includegraphics[width=1\linewidth]{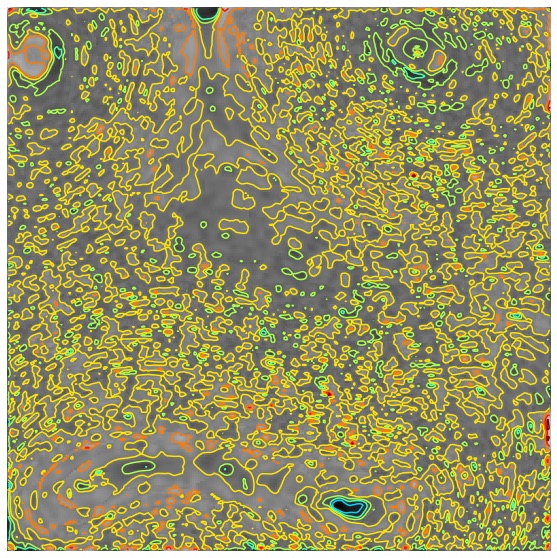}
		\caption{\centering Error SLEF-HHT (contour)}
		\label{fig:error_lefhhc}
	\end{subfigure}
	\begin{subfigure}[ht]{0.19\linewidth}
		\centering
		\includegraphics[width=1\linewidth]{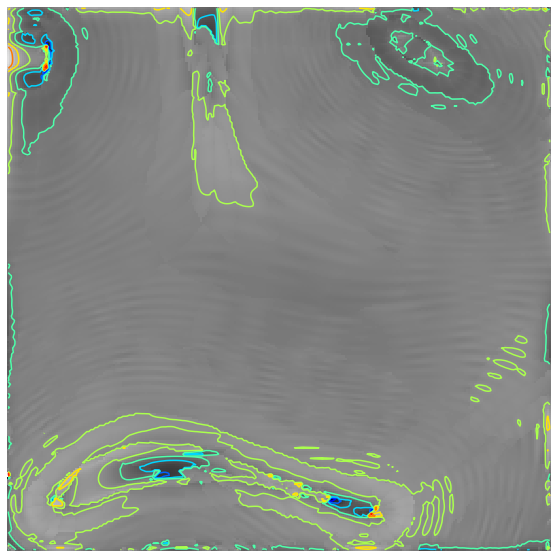}
		\caption{\centering Error SLEF-LS (contour)}
		\label{fig:error_sleflsc}
	\end{subfigure}
	\begin{subfigure}[ht]{0.19\linewidth}
		\centering
		\includegraphics[width=1\linewidth]{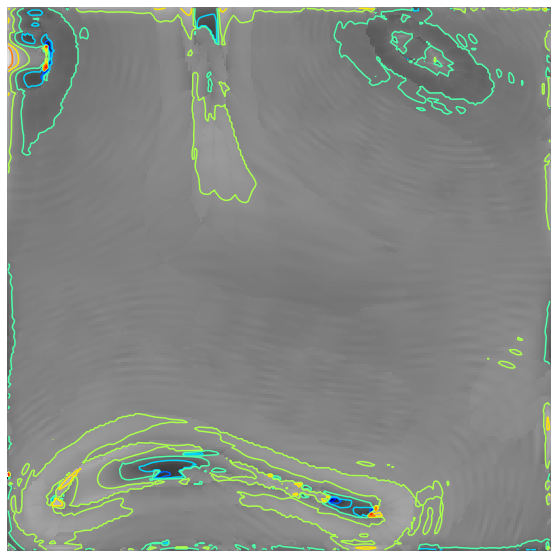}
		\caption{\centering Error SLEF-RE (contour)}
		\label{fig:error_slefrec}
	\end{subfigure}
	
	\caption{Phase estimation using LEF-GFB, SLEF-LS and SLEF-RE algorithms. (a), (b), (c) and (d) present the estimated phases. (e), (f), (g) and (h) are the wrapped error maps. (i), (j), (k) and (l) are the wrapped error maps with contour lines for visualization purposes.}
	\label{fig:phases}
\end{figure*}

In Figure \ref{fig:phases} we present the calculation of one of the synthetic patterns shown in Figure \ref{fig:var}. Figures \ref{fig:phi_lefhht} to \ref{fig:phi_slefre} present the estimated phases. Figures \ref{fig:error_lefhht} to \ref{fig:error_slefre} correspond to the wrapped error (which corresponds to the harmonics generated) between of the calculated phases with respect to the ideal phase . Figures \ref{fig:error_lefhhtc} to \ref{fig:error_slefrec} present the same wrapped error maps with contour lines for visualization purposes.

The first column corresponds to the results obtained using the LEF--HHT as proposed in \cite{liu2016simultaneous}. The second column corresponds to the LEF--HHT algorithm but using \eqref{eq:phi2} to estimate the phase map. It can be seen that the amount of errors is drastically reduced due to the use of a different equation for the phase estimation. In the third column we present the results of using the SLEF--HHT algorithm which is the use of the SLEF--RE algorithm but using the HHT as normalization method. Finally, the fourth and fifth columns correspond to the SLEF--LS and SLEF-- RE methods with GFB pre--filtering.

For this particular experiment, the error of the estimation of the step using the LEF--HHT algorithm was $\delta_{error} = 0.1194 rad$ while the error for the SLEF--LS algorithm was $\delta_{error} = 0.0567 rad$ and the SLEF--RE was $\delta_{error} = 0.0219 rad$. The SLEF--RE method with HHT pre--filtering presented an error of $\delta_{error} = 0.1092 rad$.

The MAEs of the error surfaces are $MAE_{LEF-HHT} = 1.1712$,  $MAE_{LEF-HHT-2} = 0.4722 rad$, $MAE_{SLEF-HHT} = 0.4846 rad$, $MAE_{SLEF-LS} = 0.37 rad$ and $MAE_{SLEF-RE} = 0.3569 rad$. It is important to note that the high amount of harmonics in Figure \ref{fig:error_lefhh} is due to used phase extraction equation which includes a piston term, in this case Eq. \eqref{eq:phi1}. If the phase was recovered by \eqref{eq:phi2}, the MAE would be reduced as presented in Figure \ref{fig:error_lefhhtc_mur}.
%-------------------------------------------------------------------------------------------------
\subsection{Experimental results}
\label{ssec:experimental}
%-------------------------------------------------------------------------------------------------
In order to test the capabilities of our algorithm we implemented a Polarizing Cyclic Path Interferometer (PCPI) as the one proposed in \cite{toto20174d} in its radial mode. Such system was implemented with a diode Laser with power of $400 mW$ operating at $\lambda = 532 nm$. In the arrangement, a polarizer filter is placed at an angle of $45^{\circ}$ at the entrance of the interferometer, so the incoming beam will have perpendicular and a parallel components of the same intensity. Once the light has gone through the PCPI, we placed a quarter wave plate which generates crossed circular polarization states. At the output of the PCPI, an interference pattern is observed when placing an auxiliary linear polarizer, according the next equation

\begin{equation} \label{eq:pol}
	I_k(p) = a_k(p) + b_k(p)\cos[\phi(p) + 2\Psi_k] + \eta_k(p),
\end{equation}
which is in fact the same as \eqref{eq:image}, but in this case, $\Psi_k$ corresponds to the angle of the linear polarizer. This rotation produces the phase shift between the two interferograms as $\delta_k = 2\Psi_k$.

\begin{figure}[ht]
	\centering
	\begin{subfigure}[ht]{0.32\linewidth}
		\centering
		\includegraphics[width=1\linewidth]{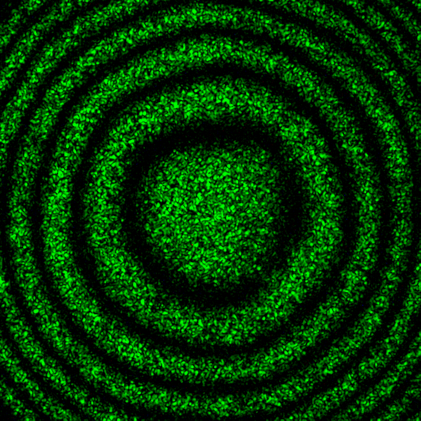}
		\caption{\centering $I_1$, $\delta=0$}
		\label{fig:I1}
	\end{subfigure}
	\begin{subfigure}[ht]{0.32\linewidth}
		\centering
		\includegraphics[width=1\linewidth]{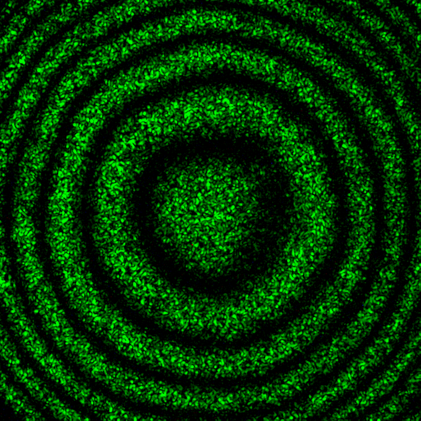}
		\caption{\centering $I_2$, $\delta=\pi/3$}
		\label{fig:I2}
	\end{subfigure}
	\begin{subfigure}[ht]{0.32\linewidth}
		\centering
		\includegraphics[width=1\linewidth]{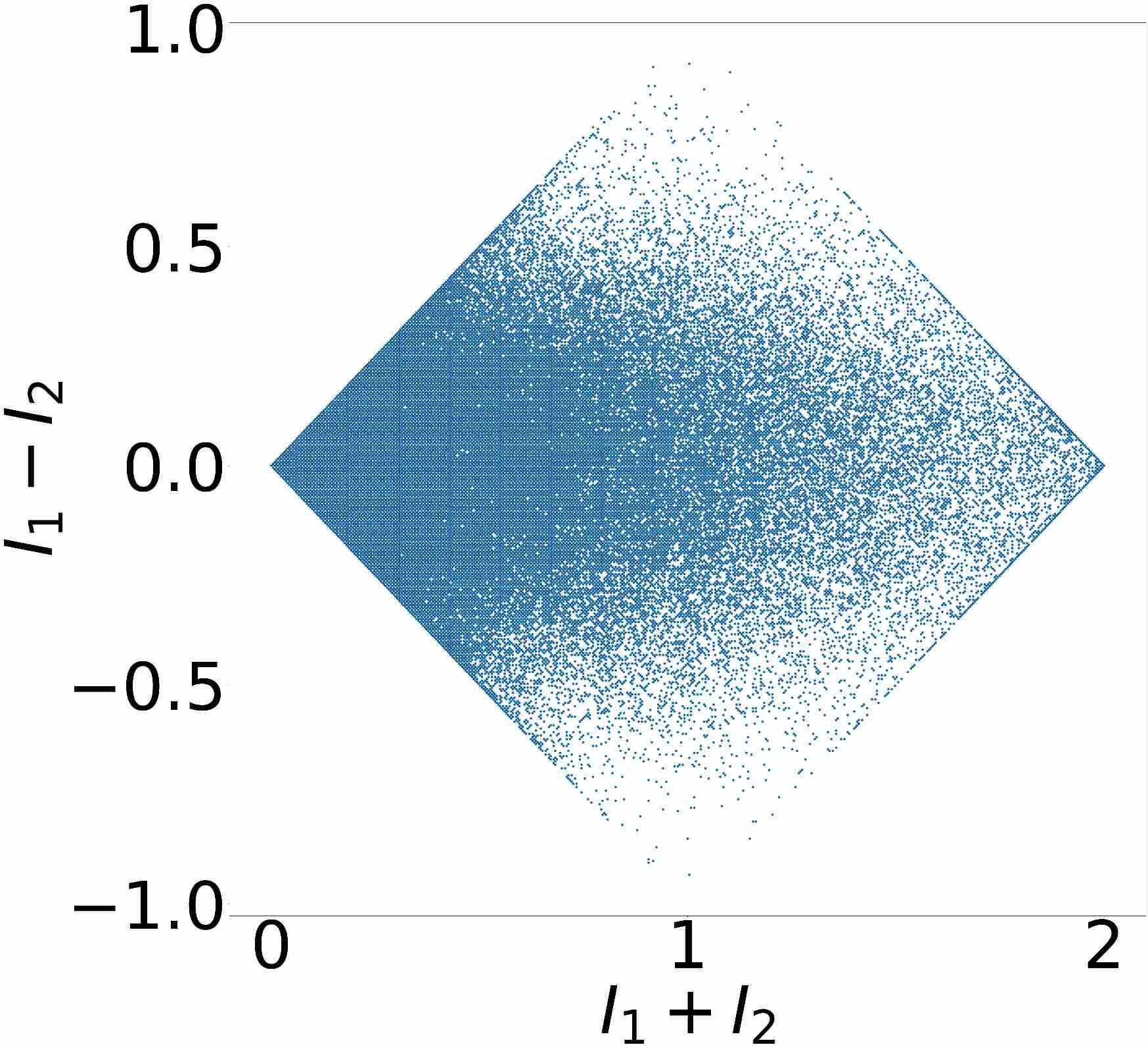}
		\caption{\centering  Lissajous Pattern}
	\end{subfigure}
	
	\begin{subfigure}[ht]{0.32\linewidth}
		\centering
		\includegraphics[width=1\linewidth]{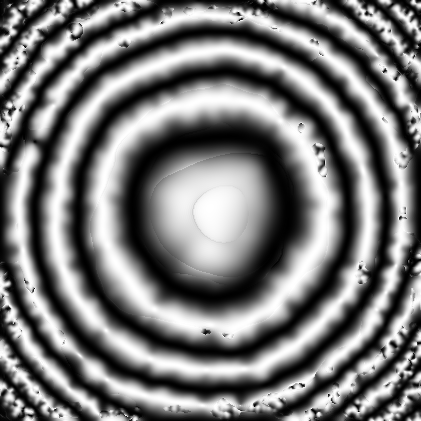}
		\caption{\centering $\hat{I}_1$, HHT pre--filtering}
		\label{fig:HHT1}
	\end{subfigure}
	\begin{subfigure}[ht]{0.32\linewidth}
		\centering
		\includegraphics[width=1\linewidth]{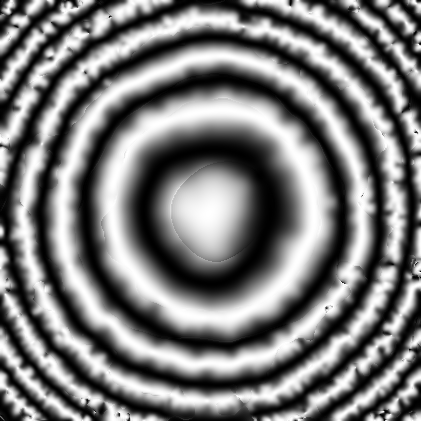}
		\caption{\centering $\hat{I}_2$, HHT pre--filtering}
		\label{fig:HHT2}
	\end{subfigure}
	\begin{subfigure}[ht]{0.32\linewidth}
		\centering
		\includegraphics[width=1\linewidth]{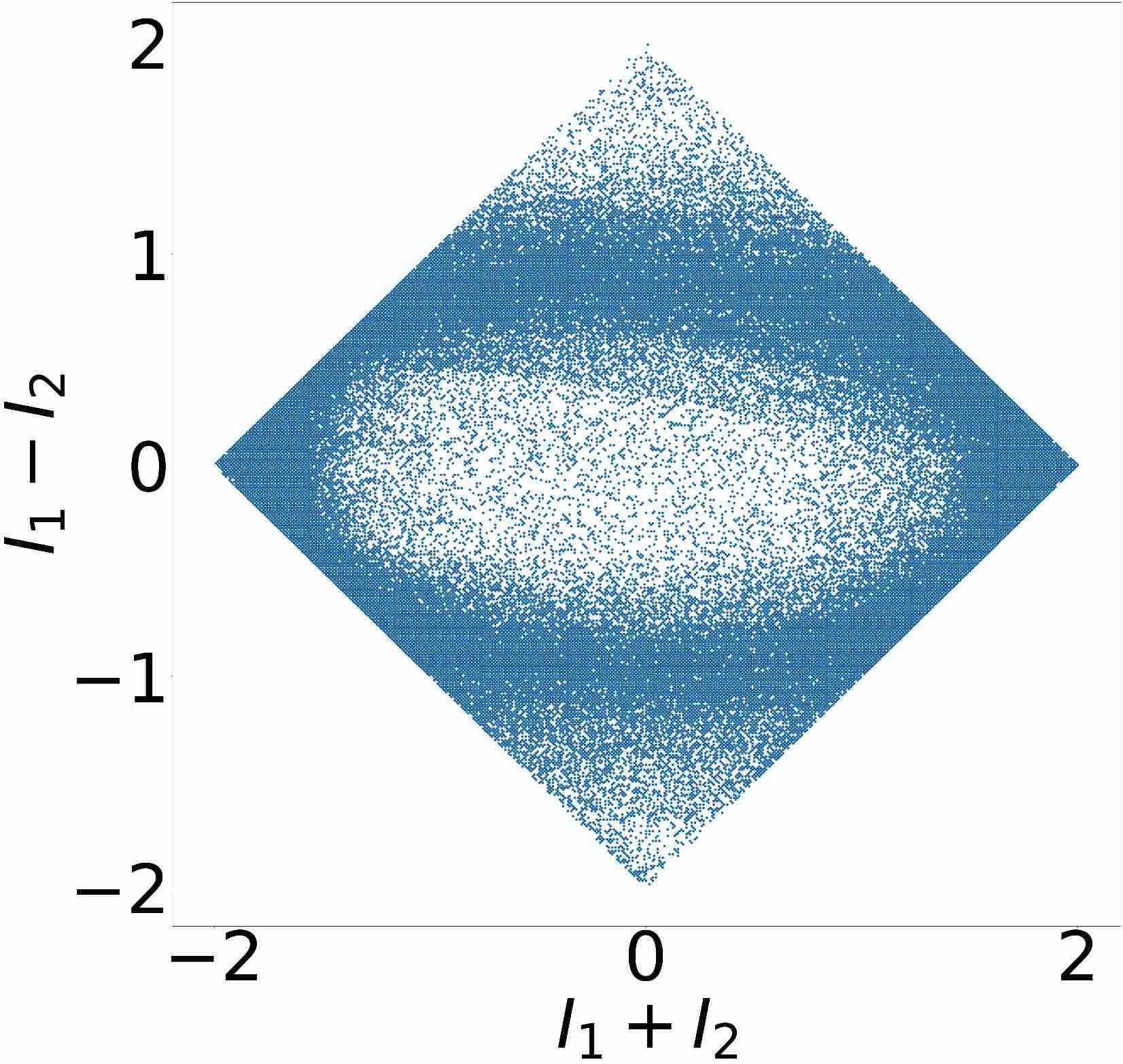}
		\caption{\centering Lissajous Pattern}
	\end{subfigure}

	\begin{subfigure}[ht]{0.32\linewidth}
		\centering
		\includegraphics[width=1\linewidth]{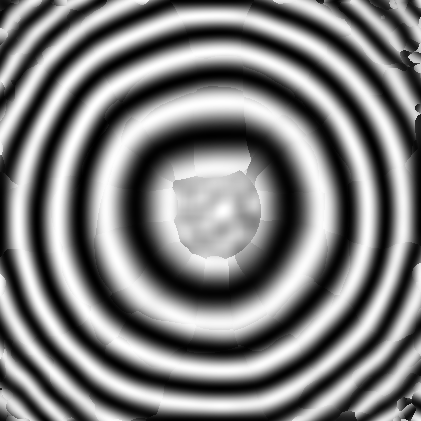}
		\caption{\centering $\hat{I}_1$, GFB pre--filtering}
		\label{fig:GB1}
	\end{subfigure}
	\begin{subfigure}[ht]{0.32\linewidth}
		\centering
		\includegraphics[width=1\linewidth]{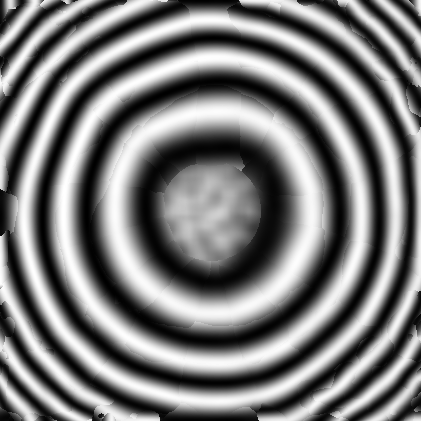}
		\caption{\centering $\hat{I}_2$, GFB pre--filtering}
		\label{fig:GB2}
	\end{subfigure}
	\begin{subfigure}[ht]{0.32\linewidth}
		\centering
		\includegraphics[width=1\linewidth]{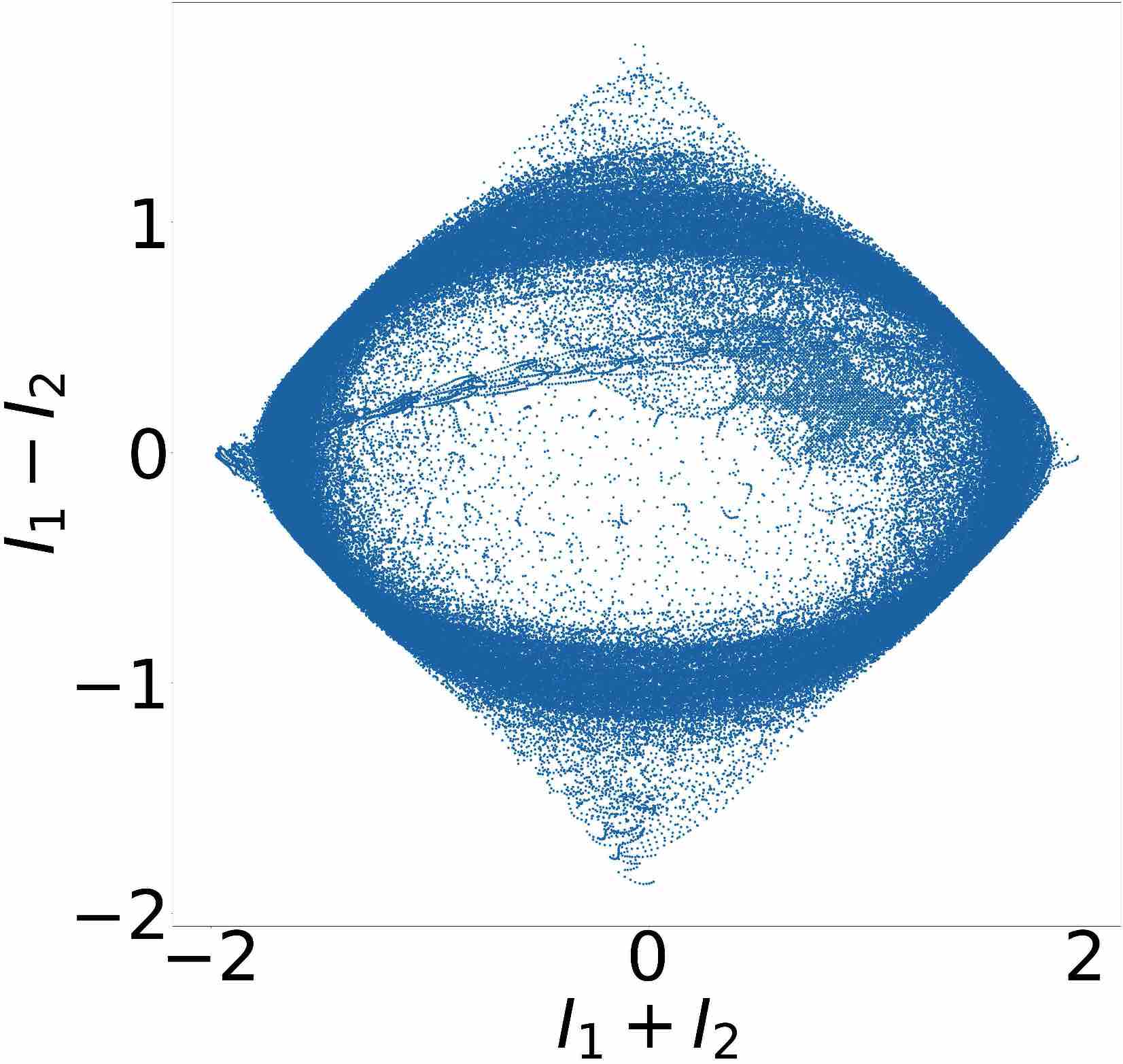}
		\caption{\centering Lissajous Pattern}
	\end{subfigure}
	\caption{Experimental results from a PCSI and filtered patterns. (a) and (b) are the experimental patterns, (c) and (d) are the patterns filtered with the HHT, (e) and (f) are the patterns filtered with GFB}
	\label{fig:exp_res}
\end{figure}

For this experiment, the auxiliar linear polarizer was mounted on a graduated rotational mount. The two interferograms were recorded with a resolution of $410\times410$ and a shift angle of $30^{\circ}$ between the captures, which means a phase step of $\delta=\pi/3$. Figures \ref{fig:I1} and \ref{fig:I2} present such patterns.

Figures \ref{fig:HHT1} and \ref{fig:HHT2} show the pre--filtered patterns using the HHT. Such result was obtained using the EFEMD algorithm with its automatic selection of modes. On the other hand, Figures \ref{fig:GB1} and \ref{fig:GB2} illustrate the normalized patterns using GFB. For these results applied a Winner--Take--All (WTA) strategy for the GFB corresponding to the periods equal to $[20, 35, 45, 55]$ and eight orientations. Since the main problem of the GFB is the detection of low frequencies, we use 

\begin{equation} \label{eq:low_freq}
	\hat{I}_k \leftarrow \alpha\hat{I}_k + (1-\alpha)\tilde{I}_k,
\end{equation}
where $\alpha=Mag/\max(Mag)$, $Mag$ is the magnitude map of the response of the GFB and $\tilde{I}_k$ is low--pass filtered version of the original image.

To prove the demonstrate the feasibility of our proposal, we present the phase obtained with the LEF--HHT, SLEF--LS and the SLEF--RE algorithms.

\begin{figure}[ht]
	\centering
	\begin{subfigure}[ht]{0.45\linewidth}
		\centering
		\includegraphics[width=1\linewidth]{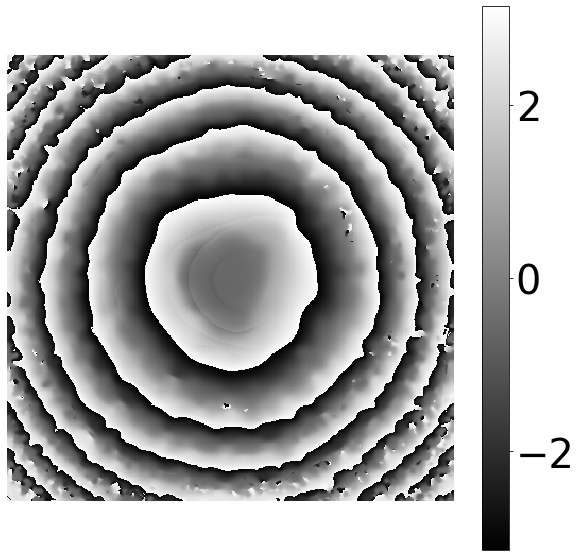}
		\caption{\centering $\phi_{LEF-HHT}$}
		\label{fig:exp_lef_hht}
	\end{subfigure}
	\begin{subfigure}[ht]{0.45\linewidth}
		\centering
		\includegraphics[width=1\linewidth]{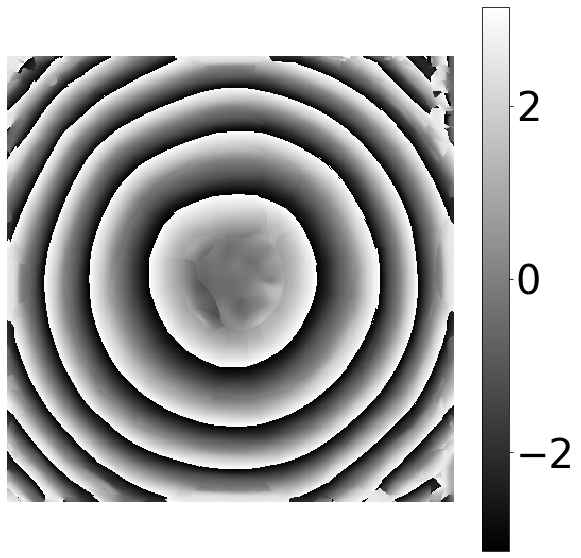}
		\caption{\centering  $\phi_{SLEF-LS}$}
		\label{fig:exp_slef_ls}
	\end{subfigure}
	
	\begin{subfigure}[ht]{0.45\linewidth}
		\centering
		\includegraphics[width=1\linewidth]{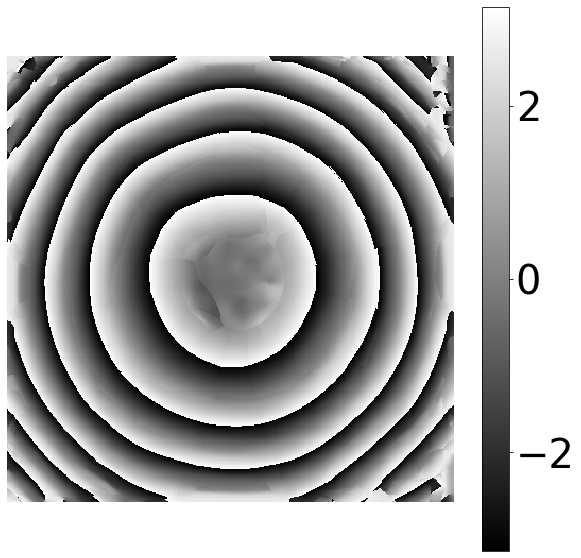}
		\caption{\centering $\phi_{SLEF-RE}$}
		\label{fig:exp_slef_re}
	\end{subfigure}
	\begin{subfigure}[ht]{0.45\linewidth}
		\centering
		\includegraphics[width=1\linewidth]{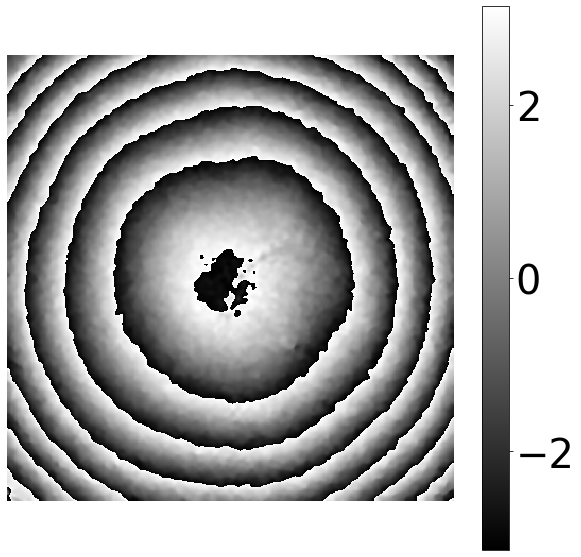}
		\caption{\centering $\phi_{4-step}$}
		\label{fig:exp_4_step}
	\end{subfigure}
	\caption{Phase maps estimated with (a) LEF--HHT, (b) SLEF--LS and (c) SLEF--RE (d) Four Steps}
	\label{fig:exp_res}
\end{figure}

Figure \ref{fig:exp_lef_hht} presents the estimated phase map using the LEF--HHT algorithm, were the pre--filtering was made with the HHT (Figures \ref{fig:HHT1} and \ref{fig:HHT2}). The estimation of the phase step by using the 5--term equation and the LS method, which gave a result of $\delta = 1.2186$, a relative error of $16.4\%$. The calculation of the phase map was made with \eqref{eq:phi1}.

Figure \ref{fig:exp_slef_ls} presents the estimated phase map using the SLEF--LS algorithm, were the pre--filtering was made with the GFB (Figures \ref{fig:GB1} and \ref{fig:GB2}). The estimation of the phase step by using the 2--term equation and the LS method (which under these conditions have the same behavior as the 5--term), which gave a result of $\delta = 1.1673$, a relative error of $11.4\%$. The calculation of the phase map was made with \eqref{eq:phi2}.

Figure \ref{fig:exp_slef_re} presents the estimated phase map using the SLEF--RE algorithm, were the pre--filtering was made with the GFB (Figures \ref{fig:GB1} and \ref{fig:GB2}). The estimation of the phase step by using the 2--term equation and the robust estimator, which gave a result of $\delta = 1.115$, a relative error of $6.4\%$. The calculation of the phase map was made with \eqref{eq:phi2}.

Figure \ref{fig:exp_res} presents the estimated phase map using the LEF-HHT, SLEF--LS and SLEF--RE.  For purposes of a qualitative comparison, we include the phase estimated with the well-known 4 steps algorithm, shown in Figure \ref{fig:exp_4_step}. The fringe patterns of this result were only filtered with a simple Gaussian filter of $\sigma=2$.

From a visual inspection, one can note that SLEF--LS and SLEF--RE algorithms present practically the same results. Meanwhile, LEF--HHT produces a result with a non smooth gray--scale transition; typical of a detunning. It is important to note the capabilities of the GFB of filtering out the noise, given that the obtained phase is clearly comparable to the 4--steps algorithm (which is more robust to noise than a two--steps method).

%-------------------------------------------------------------------------------------------------
\section{Discussions}
%-------------------------------------------------------------------------------------------------

\noindent As mentioned in section \ref{sec:LEF}, two phase-shifted interferograms can be represented as the Lissajous figure by plotting their pixel--wise corresponding intensities. Nevertheless, this algorithm requires at least that the background intensity as well as the amplitude modulation to be temporary constants \cite{farrell1992phase, liu2016simultaneous, zhang2018random, zhang2019two, meneses2015phase}. If this condition is not presented, it is required a preparation process for the interferograms in order to accomplish a constant modulation of the fringes. Preprocessing techniques such as Windowed Fourier Transform \cite{kemao2007two}, GFB \cite{rivera2016two}, the Hilbert--Huang Transform (HHT) \cite{trusiak2015two} or isotropic normalization \cite{quiroga2003isotropic} allow to eliminate such variations. Our work presents the advantage of using normalized fringe patterns since it allows us to simplify the solution of the ellipse equation from a 5--terms equation to a 2--terms form. Liu \emph{et. al} \cite{liu2016simultaneous} present a similar approach even though, they still solve a 5--terms equation. We compare our approach with theirs in order to prove the equivalency of our 2--terms SLEF--LS algorithm with the 5--term pre--filtered technique. On the other hand, the SLE--RE algorithm present more robustness to the residuals of the pre--filtering process, giving more stability and accuracy in the phase step calculation process.

%Finally, the proposals presented by Zhang \emph{et. al} do not fit with our approach: First, they use of the LEF method to calculate a first approximation of the phase distribution and then iterate the solution using a LS technique \cite{zhang2018random}. Second, they use the Gram--Schmidt orthonormalization proposed in reference \cite{vargas2012two} in order to transform the ellipse into a circle, and then, obtain the parameters to calculate the phase distribution. These proposals only use the LEF algorithm as a tool, but they are not based on the algorithm. For this reason we do not include a comparison with them.

%-------------------------------------------------------------------------------------------------
\section{Conclusions}
%-------------------------------------------------------------------------------------------------

\noindent We introduced a simplified model of the Lissajous Ellipse Fitting to calculate the phase step and phase distribution of two randomly shifted interferograms. We focused on solving the problem of obtaining the phase of interferograms with spatial--temporal dependencies on their background intensities, amplitude modulations and noise. The main advantages of use of the GFB is that the phase estimation is robust to the mentioned issues since it filters--out the noise, normalizes the amplitude and eliminates the background. Given the normalized patterns, the ellipse equation can be simplified to a two--unknowns system instead of a  five--unknowns system, we named this the SLEF algorithm. Our method consists of three stages: the preprocess the fringe patterns using a GFB, the estimation of the phase step through the estimation of the coefficients of the ellipse's equation and the calculation the phase distribution.  We remark that we can replace the GFB based preprocessing with other normalization techniques that provide the elimination of the background component, normalize the amplitude modulation and filter--out the noise. As presented, the estimation of the two terms of the ellipse equation can be done by the well--known LS method, which results are practically the same as the 5--terms. Also, we introduced a novel implementation of a robust estimator such as the Leclerc's potential in order to improve the accuracy of the phase step estimation, this is mainly caused by residuals of the filtering process. The experimental results of the calculation of our algorithms to 100 pairs of images (10 different patterns with 5 different levels of noise and 5 different phase steps) prove that SLEF--LS is equivalent to the pre--filtered 5--term LEF algorithm and the robustness to residuals of our SLEF--RE algorithm, which improves significantly the accuracy of the phase step estimation.

%-------------------------------------------------------------------------------------------------

\section*{Acknowledgements}
\noindent VHFM thanks Consejo Nacional de Ciencia y Tecnolog\'ia (Conacyt) for the provided postdoctoral grant. This research was supported in part by Conacyt, Mexico (Grant A1-S-43858) and the NVIDIA Academic program.

%-------------------------------------------------------------------------------------------------
\section*{References}

\bibliography{SLEF}

\end{document}